\documentclass[12pt,a4paper]{article}

\usepackage[centertags]{amsmath}
\usepackage{amssymb}
\usepackage{bm}
\usepackage{epsfig} 
\usepackage{amsmath} 
\usepackage{graphicx} 
\usepackage{jheppub}

\def\bea{\begin{eqnarray}}
\def\eea{\end{eqnarray}}
\def\bec{\begin{center}}
\def\ec{\end{center}}

\def\beq{\begin{equation}}
\def\eeq{\end{equation}}

\newcommand{\eq}[1]{Eq.~(\ref{#1})}
\newcommand{\eqs}[2]{Eqs.~(\ref{#1}) and (\ref{#2})}

\newcommand{\eV}{\mathinner{\mathrm{eV}}}

\newcommand{\MeV}{\mathinner{\mathrm{MeV}}}
\newcommand{\GeV}{\mathinner{\mathrm{GeV}}}
\newcommand{\TeV}{\mathinner{\mathrm{TeV}}}

\newcommand{\planck}{\mathrm{P}}
\newcommand{\nlsp}{\chi}
\newcommand{\axino}{{\tilde{a}}}

\title{Cosmological moduli problem in large volume scenario and thermal inflation}

\author[a]{Kiwoon Choi,} 
\author[b]{Wan-Il Park,}
\author[c]{and Chang Sub Shin}

\affiliation[a]{Department of Physics, KAIST, \\ Daejeon 305-701, Korea}
\affiliation[b]{School of Physics, KIAS, \\ Seoul 130-722, Korea}
\affiliation[c]{APCTP, Pohang, Gyeongbuk 790-784, Korea}

\emailAdd{kchoi@kaist.ac.kr}
\emailAdd{wipark@kias.re.kr}
\emailAdd{csshin@apctp.org}

\abstract{
We show that in a large volume scenario of type IIB string or F-theory compactifications, single thermal inflation provides only a partial solution to the cosmological problem of the light volume modulus.  We then clarify the conditions for double thermal inflation, being a simple extension of the usual single thermal inflation scenario, to solve the cosmological moduli problem in the case of relatively light moduli masses.  Using a specific example, we demonstrate that double thermal inflation can be realized in large volume scenario in a natural manner, and the problem of the light volume modulus can be solved for the whole relevant mass range.  We also find that right amount of baryon asymmetry and dark matter can be obtained via a late-time Affleck-Dine mechanism and the decays of the visible sector NLSP to flatino LSP.
}

\keywords{  }

\begin{document}

\maketitle

\section{Introduction}

One of the most attractive features of low energy supersymmetry (SUSY) is the unification of gauge couplings at the scale $M_{\rm GUT} \sim 10^{16} \GeV$.
The GUT scale  is two orders of magnitude smaller than the Planck scale, and this hierarchy may be understood within a higher dimensional setup of string theory, in which gravity lives in a relatively large bulk spacetime, while GUT degrees of freedom are confined on branes \cite{Witten:1996mz}.
Such scenario then requires a moduli stabilization scheme which would  stabilize the overall volume modulus at a large value in string unit.

An attractive scheme realizing large compactification volume is the large volume scenario (LVS) of \cite{lvs}, proposed in the context of type IIB string theory.
This scheme  involves two K\"ahler moduli, $T_b$ and $T_s$, where $\tau_b={\rm Re}(T_b)$ corresponds to a 4-cycle volume modulus determining
the bulk Calabi-Yau volume  as ${\cal V}\simeq \tau_b^{3/2}$, and $\tau_s={\rm Re}(T_s)$ 
describes a small 4-cycle volume supporting 
a nonperturbative superpotential of the form $W_{\rm np} \sim e^{-aT_s}$. 
The interplay between $W_{\rm np}$  and 
a perturbative correction to the K\"ahler potential suppressed by 
$1/\tau_b^{3/2}$ determines the VEV of $\tau_b$ as \bea
\tau_b^{3/2} \sim e^{a\tau_s},\eea which can be exponentially large for a moderately large
value of  $a\tau_s$.
In LVS,  the hierarchy between the Planck scale and the GUT scale is given 
by  \cite{Conlon:2009xf}
 \bea M_{\rm GUT}/M_{\planck} \sim 1/\tau_b^{1/2},\eea so $\tau_b = \mathcal{O}(10^4)$ results in the correct hierarchy between the GUT scale and the Planck scale.

The large VEV of the volume modulus implies that its scalar potential is relatively flat (at least near the minimum), so the volume modulus is relatively light.
Specficially, the mass of the large volume modulus in LVS is given by 
\bea
m_{\tau_b} 
\sim \frac{1}{\sqrt{a\tau_s}} \frac{m_{3/2}}{ \tau_b^{3/4}}
\sim \frac{m_{3/2}}{\sqrt{\ln (M_{\planck}/m_{3/2})}} \left(\frac{M_{\rm GUT}}{M_{\planck}}\right)^{3/2} 
\eea
which would be in the range of cosmological care unless the gravitino mass $m_{3/2}$ is heavy enough to ensure
$m_{\tau_b}\gtrsim 100$ TeV. 
Cosmologically, huge amount of moduli can be produced in the form of a coherent oscillation after primordial inflation, so in general light moduli cause disasters due to their too late decays. 
Depending on their life-times, they could destruct light elements formed at BBN, produce too much $X$ ($\gamma$)-rays, distort CMBR or provide too much relic density \cite{Coughlan:1983ci,Choi:1998dw}.
One way out of this cosmological moduli problem is raising-up of the moduli masses, so that moduli can decay before BBN.
For this,  the volume modulus should be heavier than about $100 \TeV$ as usual moduli should.
For $\tau_b ={\cal O}(10^4)$, such heavy volume modulus would require $m_{3/2} \gtrsim 10^9$ GeV,  
and therefore a careful sequestering of gravity mediated SUSY breaking if one wishes to realize low energy SUSY in the visible sector with soft SUSY breaking masses  $m_{\rm soft}\sim 1$ TeV \cite{Blumenhagen:2009gk}.

On the other hand, considering various possible higher order corrections,
either in $\alpha^\prime$ or in the string coupling,  it is likely to be difficult to realize a sequestering which would allow
$m_{\rm soft}/m_{3/2}\sim 10^{-6}$ \cite{Conlon:2010ji,Choi:2010gm,Shin:2011uk}.
In case that  an enough sequestering is not achieved, 
the large volume modulus becomes relatively light, and then the most compelling solution to the cosmological moduli problem is a late thermal inflation \cite{Lyth:1995hj}.
As primordial inflation does, thermal inflation can wipe out pre-existing particles, for example, moduli.
However it works in a limited sense, since energy density of thermal inflation itself becomes a source of moduli reproduction.
Depending on the mass of moduli, such a reproduction could be fatal, and the large volume scenario could be in cosmological trouble. 

In this paper, we study the cosmological moduli problem of LVS for the gravitino mass range
\beq \label{m32-band}
10^2\ {\rm GeV} \lesssim m_{3/2}\lesssim 10^9\ {\rm GeV},
\eeq
which corresponds to the volume  modulus mass range 
\beq \label{mtaub-band}
10^{-2} \GeV \lesssim m_{\tau_b} \lesssim 10^5 \GeV
\eeq
for $\tau_b = \mathcal{O}(10^4)$.
We show that, if inflaton decays before the large volume modulus starts to oscillate, single thermal inflation can solve the problem only for a limited range of $m_{\tau_b}$. 
Particularly, the volume modulus mass range $m_{\tau_b} = \mathcal{O}(10^{-2} - 1) \GeV$, which would be the case if gravity mediated SUSY breaking is not 
sequestered well,  is difficult to be cosmologically viable
with single thermal inflation only. 
On the other hand, if inflaton decays just before thermal inflation begins, the abundance of moduli can be minimized, so the whole range of $m_{\tau_b}$ can be viable with a single-step thermal inflation.
However, such a very late decay of inflaton is not typical, so we consider a double-step thermal inflation as an alternative solution, and show that the problem can be solved in a natural manner.
It turns out that correct amounts of baryon asymmetry and dark matter can be obtained also via a late-time Affleck-Dine leptogenesis \cite{Stewart:1996ai,Jeong:2004hy,Felder:2007iz,Kim:2008yu,Choi:2009qd} and the decay of the visible sector NLSP to flatino LSP.

This paper is organized as follows.
In section 2, we briefly review LVS and the properties of moduli.
In section 3, cosmological moduli problem in LVS is discussed.
In section 4, it is argued that single thermal inflation is not enough to solve the cosmological moduli problem for certain range of the volume modulus mass. 
We then clarify the conditions for a double-step thermal inflation  
to be realized, while solving the cosmological moduli problem for the entire modulus mass range
in consideration.   
In section 5, considering a specific model for the mediation of SUSY breaking,
which involves  $D$-term mediation associated with anomalous $U(1)$
gauge symmetry  and also the conventional gauge mediation, we demonstrate that a double-step thermal inflation can be realized in a natural manner and solve the cosmological  problem of the light volume modulus. 
We also show that the model can produce  correct amounts of
baryon asymmetry and dark matter too.
Section 6 is the conclusion.

\section{Large volume scenario and properties of moduli}

In this section, we review briefly the large volume compactification and properties of moduli involved.

The basic properties of LVS models can be viewed from a simple model in 4D 
supergravity (SUGRA) framework. 
The model is characterized by the following K\"ahler potential and superpotential \cite{lvs}
\bea\label{moduli_model}
K &=& -3\ln \tau_b + \frac{2(\tau_s^{3/2}-\xi_{\alpha'})}{\tau_b^{3/2}}
+ {\cal O}\left(\frac{1}{\tau_b^{3}}\right) \nonumber
\\
W &=& W_0 + A_s e^{-a T_s}
\eea
where $\tau_I={\rm Re}(T_I)$ ($I=b,s$) are the K\"ahler moduli 
determining the size of the Calabi-Yau (CY) volume 
as ${\cal V} = \tau_b^{3/2}-\tau_s^{3/2}$, and
$K$ is expanded in powers of $1/\tau_b$.
The $\xi$-term is the stringy $\alpha'$ correction and given by $\xi_{\alpha'} = \zeta(3) \chi(M) / \left[ 2 g_s^{3/2} \left( 2 \pi \right)^3 \right]$ with $\chi(M)$ and $g_s$ being the Euler number of the CY manifold and string coupling, respectively. $W_0$ is 
the tree-level flux superpotential from the stabilization of 
the dilaton and complex structure moduli. 
$A_s$, the flux dependent constant,  
can be taken to be real value by  shift transformation of $T_s$: 
$T_s\rightarrow T_s + i\alpha$. Then,
one finds that the SUGRA scalar potential at large volume limit is given by
\beq
V = \frac{2\sqrt{2}}{3} \frac{\left( a A_s \right)^2 \tau_s^{1/2} e^{- 2 a \tau_s}}{\tau_b^{3/2}} -
\left( \frac{2a A_s W_0^* \tau_s e^{- a T_s}}{\tau_b^3} + h.c. \right)
+ \frac{3 \xi_{\alpha'}}{2} \frac{|W_0|^2}{\tau_b^{9/2}}
\eeq 
This potential has a SUSY-breaking AdS minimum at 
\beq
\tau_b^{3/2} \simeq \Big(\frac{3\sqrt{2}|W_0|}{aA}\Big) \sqrt{\tau_s} e^{ a \tau_s},
 \quad \tau_s^{3/2} \simeq \xi_{\alpha'}
\eeq
Since it is known that possible uplift process to get dS vacuum does not change the vacuum properties related to our argument, we do not care about that in this paper.
The masses of the moduli are given by
\bea \label{masses}
m_{\phi_s} &\simeq & m_{a_s} \simeq m_{\psi_s}
\sim  m_{3/2}\ln \frac{M_\planck}{m_{3/2}}, 
\nonumber\\
m_{\phi_b} &\sim& 
\frac{m_{3/2}}{\tau_b^{3/4}}\sqrt{\frac{1}{\ln (M_\planck/m_{3/2})}},\quad 
m_{a_b} \simeq 0,\quad m_{\psi_b}=m_{3/2},
\eea
where $M_{\planck}=2.4 \times 10^{18} \GeV$ is the reduced Planck scale, 
\bea \label{canonical}
&&\phi_s \simeq M_\planck \tau_s^{3/4}/\langle\tau_b^{3/4}\rangle,\  
a_s\simeq M_\planck{\rm Im}(T_s)/\langle\tau_s^{1/4}\tau_b^{3/4}\rangle,\nonumber\\
&&\phi_b \simeq M_\planck\ln\tau_b,\hskip 1.25cm a_b\simeq 
M_\planck{\rm Im}(T_b)/\langle\tau_b\rangle\eea
 are the canonically normalized scalar fields of $T_s$ and $T_b$, respectively, 
 $\psi_s$ and $\psi_b$ are their  fermionic superpartners, 
$m_{3/2} =e^{K/2}|W| \simeq |W_0| / \tau_b^{3/2}$
is the gravitino mass.
A generic feature of moduli masses in LVS is that the mass of the large volume modulus is suppressed by a power of $1/\tau_b$ compared to $m_{3/2}$, reflecting the fact that  the no-scale structure 
of the scalar potential is approximately preserved in the large volume limit. 
The enhancement or suppression factor, the power of $\ln(M_{\planck}/m_{3/2})$, in the mass spectrum (\ref{masses}) is a consequence of the non-perturbative corrections which have a crucial role in moduli stabilization.

In order to know the real values of the moduli masses,
we have to specify the vacuum value of $\tau_b$ and the gravitino mass. 
In the context of  local GUT models, for universal tree-level gauge kinetic functions $f_a$,
the running gauge coupling constants  are unified at scale $M_{\rm GUT}\sim M_{\planck}/\tau_b^{1/2}$ \cite{Kaplunovsky:1994fg,Conlon:2009xf}.  
Therefore, the hierarchy between the GUT  and Planck  scales can be naturally 
obtained for $\tau_b\sim 10^3-10^5$ as
\bea
\tau_b^{1/2}\sim \frac{M_{\rm GUT}}{M_{\planck}}\sim 10^2.
\eea
The gravitino mass is an order parameter of SUSY breaking, and its value 
is related with soft SUSY breaking masses for the visible sector, $m_{\rm soft}= {\cal O}(1\,{\rm TeV})$,
by SUSY breaking mediation mechanism.
In the mixed  $D$-term-gauge mediation which will be discussed in our paper, $m_{\rm soft}$ is 
just of the order of $m_{3/2}$. In other mediation mechanism, such as   
small volume modulus mediation $(m_{\rm soft}\sim m_{3/2}/8\pi^2$)
\cite{lvs} whose mass hierarchy is similar to that of \cite{mirage1} as a general result for the SUSY breaking moduli stabilized  by 
non-perturbative superpotential, and $U(1)_A$ threshold corrections 
($m_{\rm soft}\sim m_{3/2}/\sqrt{8\pi^2}$) \cite{Shin:2011uk} in which the soft mass squared is generated at one-loop order due to 
the SUSY breaking mass spectrum of the $U(1)_A$ vector superfield, 
there can be hierarchy between $m_{3/2}$ and soft terms.
On the other hand, as discussed in \cite{Blumenhagen:2009gk}, 
if the visible sector is far from the small 4-cycle governing by $\tau_s$, 
and  the possible loop corrections are quite suppressed,
then tree-level gravity mediation induces the soft parameters much 
suppressed compared to the gravitino mass as $m_{\rm soft} \sim m_{3/2}/\tau_b^{3/2}$.
\footnote{In this type of model, the typical sfermion mass $m_{\varphi_i}$ 
is of order of $m_{\phi_b}\sim m_{3/2}/\tau_b^{3/4}$, while 
the gaugino mass $M_a$ is of ${\cal O}(m_{3/2}/\tau_b^{3/2})$.
In order for $m_{\varphi_i}$ to be the same order of $M_a$, 
we need further sequestering as $e^{-K/3} Z_i \sim 1 + {\cal O}(1/\tau_b^3)$, 
where $Z_i$ is the matter K\"ahler metric, or  $m_{\varphi_i}^2$ is given by $(\partial V/\partial\phi_b)/
M_{\planck} + {\cal O}(m_{3/2}^2/\tau_b^{3})$ 
so that $\langle m_{\varphi_i}^2\rangle = {\cal O}(m_{3/2}^2/\tau_b^3)$.}
Thus, here we allow the gravitino mass in the range of  
\bea
10^2\,{\rm GeV}\lesssim m_{3/2}\lesssim 10^6\,{\rm TeV},
\eea
that gives 
\bea
1 \TeV \lesssim &m_{\phi_s}& \lesssim 10^7 \TeV,
\\ \label{m-phib}
10\,{\rm MeV}\lesssim &m_{\phi_b}& \lesssim 10^2\,{\rm TeV}.
\eea

The life-time of a modulus is determined by interactions with its superpartners and 
ordinary matter fields localized in small cycles. In the latter case, 
most important parts are the K\"ahler metric and the 
gauge kinetic function for the visible sector 
\bea \label{mat_action}
 K_{\rm vis}&=&  Z_i \Phi_i^*\Phi_i +\cdots, \nonumber\\
 f_a &=& k_a T_v+\cdots.
\eea
where $\Phi_i$s are the matter chiral superfields, 
$T_v$ is the K\"ahler modulus determining the 
visible sector 4-cycle volume, and ($\cdots$) denotes higher order terms.
$T_v$ can not be identified as $T_b$, if not, it provides 
too small gauge coupling constants. This is the phenomenological 
reason why the matter fields should be
localized in a small cycle. In this case, the functional form of the K\"aher metric is given 
by $ Z_i \simeq {\cal Y}_j(\tau_s, \tau_v)/\tau_b$. 
Then, it is straightforward to calculate the coefficients of the following  interaction terms 
between the large volume modulus and the visible sector fields:
\bea\label{moduli-matter}
&&\frac{\delta \phi_b}{M_{\planck}}\left\{\beta_{a} F^{a\mu\nu}F^a_{\mu\nu} 
+ \tilde c_i m_{\varphi_i}^2 |\varphi_i|^2 +(\tilde c_{ij}b_{\varphi_{ij}} \varphi_i\varphi_j + h.c.)
\frac{}{}\right.\nonumber\\&&\hskip 1cm\left.
+\,\Big(c_{a} M_a \lambda^a\lambda^a + 
c_i m_{\psi_i} \psi_i\psi_i^c+ h.c.\Big) \right\}.
\eea
The visible-sector gauge field strength is denoted by $F^{a\mu\nu}$, 
$\lambda^a$ is the superpartner of the gauge field, the gaugino.
$\varphi_i$ and $\psi_i$ are the scalar and fermion component of $\Phi_i$, 
respectively.
$m_{\varphi_i}^2$ and $b_{\varphi_{ij}}$ are quadratic mass 
parameters for sfermions 
coming from SUSY breaking and supersymmetric contributions, respectively.
$M_a$ is the gaugino mass,  $m_{\psi_i}$ is
the fermion mass for $\psi_i+\psi_i^c$.
The model dependent coefficients are generically
\bea
\beta_a={\cal O}\Big(\frac{1}{8\pi^2}\Big),\ 
\tilde c_i\sim 1\ {\rm or }\ \frac{m_{\phi_b}^2}{m_{\varphi_i}^2}
,\ \tilde c_{ij}\sim 1\ {\rm or }\ \frac{m_{\phi_b}^2}{b_{\varphi_{ij}}},\
c_a\sim c_i\sim 1.
\eea
Let us briefly discuss the origin of each terms. 
At a tree-level, the interactions between $\phi_b$ and gauge fields are rather
suppressed, because $f_a$ in Eq.~(\ref{mat_action}) does not have 
$T_b$ dependence. However, due to the Konishi and the super-Weyl anomaly, 
three point interactions are generated at the one-loop level \cite{Kaplunovsky:1994fg}.
Thus $\beta_a$ is generically ${\cal O}(1/8\pi^2)$.
$\tilde c_i$, $\tilde c_{ij}$ and $c_a$ are mostly originated from the 
soft SUSY breaking mass terms that depend on $\phi_b$ as 
$m_{\rm soft}\sim  M_\planck/\tau_b^{n_b}\sim M_{\planck} \exp(- n_b \phi_b/M_{\planck})$, 
where $n_b$ is the constant
determined by SUSY breaking and its mediation mechanism, so generically 
they are of order unity.
In some cases,  
there are terms like $(\partial V/\partial\phi_b)/M_{\planck}$ 
in $m_{\varphi_i}^2$ and $b_{\varphi_{ij}}$,
which are vanishing by equations of motion so that the contribution to 
the soft masses are zero. However, the interation between the modulus 
and the scalar fields are extracted as $(\partial^2 V/\partial^2\phi_b)\,
\delta \phi_b\, |\varphi_i|^2/M_{\planck}\sim m_{\phi_b}^2 \delta \phi_b |\varphi_i|^2/
M_{\planck}$. As a result, $\tilde c_i$, and $\tilde c_{ij}$ can be enhanced by a factor of 
$m_{\phi_b}^2/m_{\rm soft}^2$ for $m_{\phi_b}\gg m_{\rm soft}$.\footnote{
It can be shown that we can take a field basis such that derivative interactions between 
$\phi_b$ and ($\varphi_i,\, \psi_i$) are absent. 
Then, all interactions are obtained from the potential terms. 
This basis is good in the sense that we do not need to care about 
cancelations between the interactions 
obtained from the kinetic part and potential part.}
$c_i$ depends on the origin of the fermion mass term. 
Through the Higgs bilinear $\mu$-parameter and soft masses,
the Higgs can interact with $\phi_b$. Due to the non-zero vacuum value of the Higgs fields,  
there is a mixing between the Higgs and $\phi_b$ in the mass eigenbasis.
Then, the SM fermions can interact with $\phi_b$ through the modulus-Higgs mixing.
Higgsinos can directly interact with $\phi_b$ via $\mu H_u H_d$ term in the superpotential
or from the K\"ahler potential by Giudice-Masiero mechanism \cite{Giudice:1988yz}.
In addition, the large volume modulus can decay to axion $a_b$, the complex counter part of the modulus, with a sizable branching fraction via kinetic term.

All those interactions are suppressed by Planck scale with additional suppression factors depending on channels.
The decay rates of $\phi_b$ to daughter particles are given by 
\bea\label{decay mode}
&&\Gamma_{\phi_b\rightarrow 2a_b} \simeq 
\frac{1}{48\pi} \frac{m_{\phi_b}^3}{M_{\planck}^2},
\ \  \, \quad\frac{\Gamma_{\phi_b\rightarrow 2A^a_\mu}}
{\Gamma_{\phi_b\rightarrow 2a_b}} 
\simeq  \frac{(8\pi^2\beta_{a})^2}{(8\pi^2)^2},\
\frac{\Gamma_{\phi_b\rightarrow \varphi_i\varphi_i^*}}{\Gamma_{\phi_b\rightarrow 2 a_b}} 
\simeq\tilde c_i^2\Big(\frac{ m_{\varphi_i}^2}{m_{\phi_b}^2}\Big)^2,\ \nonumber\\
&&
\frac{\Gamma_{\phi_b\rightarrow \varphi_i\varphi_j}}
{\Gamma_{\phi_b\rightarrow 2a_b}} \simeq
\tilde c_{ij}^2\Big(\frac{ b_{\varphi_{ij}}}{m_{\phi_b}^2}\Big)^2, \
\frac{\Gamma_{\phi_b\rightarrow 2\lambda^a}}{\Gamma_{\phi_b\rightarrow 2a_b}} 
\simeq c_a^2\Big(\frac{M_a^2}{m_{\phi_b}^2}\Big),\
\frac{\Gamma_{\phi_b\rightarrow \psi_i\psi_i^c}}{\Gamma_{\phi_b\rightarrow 2 a_b} }
\simeq c_i^2\Big(\frac{ m_{\psi_i}^2}{m_{\phi_b}^2}\Big),
\eea
provided  the decay channels are kinematically allowed.  
Note that decays to matter fields is at most comparable to the axion channel though it depends on models.
Note also that the branching fraction of the large volume modulus to photons is suppressed by $\mathcal{O}(10^{-4})$ relative to that of axion channel.

On the other hand, the strength of interactions between the small volume modulus and 
the matter fields is much more enhanced compared 
to that for the large volume modulus. 
Because, as in Eq.~(\ref{canonical}), $\delta \tau_s \sim \delta \phi_s/M_{\rm st.}$ with $M_{\rm st}\simeq  M_{\planck}/\tau_b^{3/4}$ being the string scale in large volume scenario, 
the suppression scale of 
the interaction between $\phi_s$ and matters is 
not Planck scale, but string scale \cite{Conlon:2007gk}. 
As a result,
in case that $T_v$ and $T_s$ have a sizable mixing at the tree-level,
the small volume modulus mostly decays to the gauge boson pair as
\bea
\Gamma_{\phi_s} &\simeq & \Gamma_{a_s}\simeq
\frac{\gamma_{s \to A_\mu^a A_\mu^a}}{64 \pi } 
\frac{m_{\phi_s}^3}{M_{\rm st}^2},
\eea 
where $\gamma_{s \to A_\mu^a A_\mu^a}$ is a numerical coefficient of $ \mathcal{O}(10)$ taking the number of allowed channels into account.

\section{Cosmological moduli problem in LVS}
\begin{figure}[h] 
\centering
\includegraphics[width=0.8\textwidth]{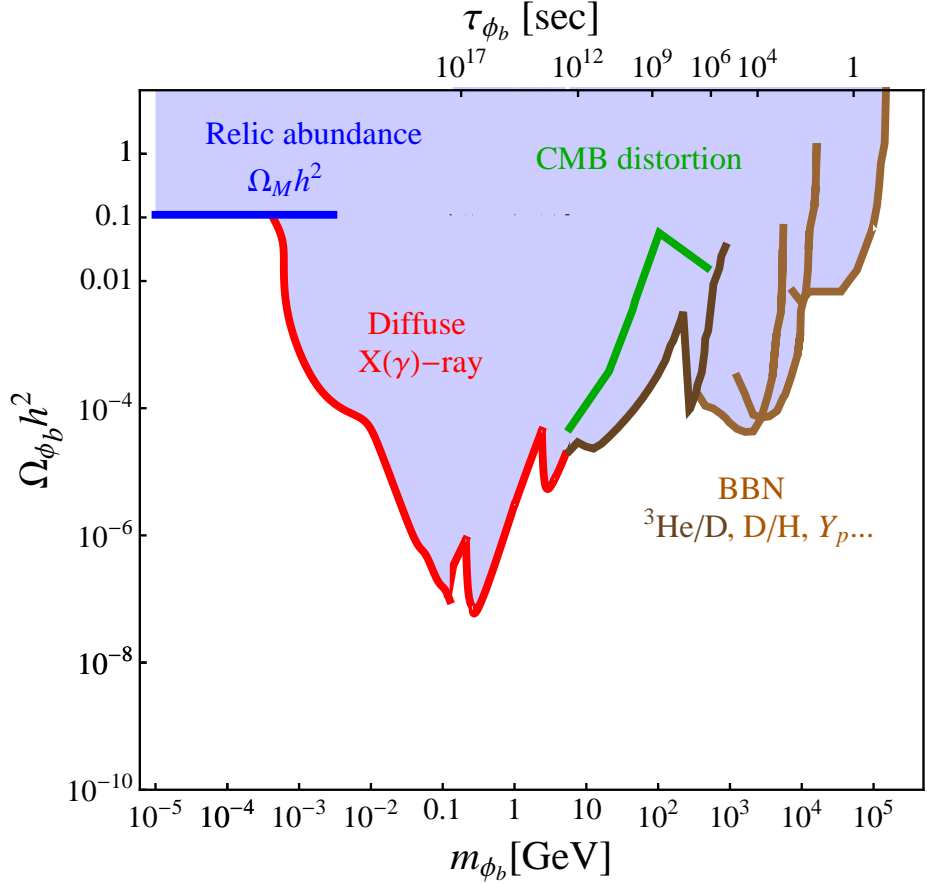} 
\caption{ Constraints on $\Omega_{\phi_b}h^2$ 
before $\phi_b$ decays are presented for a wide range of the modulus mass.
Allowed region is below the lines obtained 
from the present dark matter density (blue), 
diffuse X($\gamma$)-ray (red), CMB distortion (green), 
and light elements abundance (dark brown and brown).  
From the conservative point of view,  we assume 
$\tilde c_{H_uH_d} = {\cal O}(m_{\phi_b}^2/b)$, 
so that the branching ratio for $\phi_b\rightarrow$ SM 
is sizable in the mass range of $m_{\phi_b} > 2m_{\rm Higgs}$.}
\label{fig:mod_problem}
\end{figure}

The abundance of a modulus is highly constrained, depending on its life-time. 
The small volume modulus decays dominantly to SM particles well before BBN epoch, hence it is harmless as long as right amounts of baryon number asymmetry and dark matter can be obtained.
In case of the large volume modulus, the life-time
can be written as
\bea
\tau_{\phi_b} = 0.6\sec\, {\rm Br}_{\phi_b \to a_ba_b} 
\left(\frac{100\,{\rm TeV}}{m_{\phi_b}}\right)^3
\eea
where ${\rm Br}_{\phi_b \to a_ba_b}$ is the branching fraction to 
large volume axions and always sizable for all mass range of $\phi_b$.
For $m_{\phi_b} \lesssim 100 \TeV$, the large volume modulus decays around or after BBN epoch.
In this case, as listed below, there are various constraints depending on the life-time, $\tau_{\phi_b}$.
\begin{itemize}
\item $\tau_{\phi_b} \lesssim 0.1 \sec$: 
The relic abundance is restricted by the effective number of the extra relativistic degrees of freedom, $\Delta N_{\rm eff}$, to which the primordial abundance of $^4$He, D is highly sensitive.
Contribution from moduli decay is
\beq
\Delta N_{\rm eff} = \frac{\rho_{a_b}}{\rho_{\nu}} = \frac{1 - {\rm Br}(\phi_b \to {\rm SM})}{{\rm Br}(\phi_b \to {\rm SM}) \times g_{\nu} / g_*(T_{\rm d, \phi_b})}
\eeq
where $g_\nu = 7/4$ and $g_*(T_{\rm d, \phi_b})$ are the relativistic degrees of freedom of a left-handed neutrino and standard model particles (including neutrinos), respectively.
The observed value, $\Delta N_{\rm eff}< 1$ ($1\sigma$) from BBN \cite{Mangano:2011ar} 
and $\Delta N_{\rm eff}= 0.85^{+0.39}_{-0.56}$ ($1\sigma$) from BBN+CMB+LSS \cite{Hamann:2011ge}, poses a constraint,  ${\rm Br}_{\rm SM} \gtrsim 0.9$.
Such a sizable branching fraction can be achieved if $\tilde c_i\, (\tilde c_{ij})\sim m_{\phi_b}^2/m_{\rm soft}^2$ \cite{Cicoli:2012aq}. 
Otherwise, for $\tilde c_i (\tilde c_{ij})\sim 1$, ${\rm Br}_{\rm SM}\ll 1$, 
and moduli should be a sub-dominant energy component of the universe when they decay.
Actually, 
the constraint from the extra radiation contribution to the SM radiation is applicable for the moduli decaying in a present Hubble time though $g_\nu$ has to be replaced to $g_\nu \times (4/11)^{4/3}$ for $\tau_{\phi_b} \gtrsim 1 \sec$.
However, the constraint is weaker than others described below. 
\item $0.1\sec \lesssim \tau_{\phi_b} \lesssim 10^{12} \sec$: 
The direct interaction between decay products 
of $\phi_b$ and the background  light elements causes significant changes in the abundances of light elements from the predictions by standard BBN \cite{Jedamzik:2004er}. 
If $\tau_{\phi_b} \gtrsim 10^6 \sec$, it also affects the CMB spectrum as a deviation from the Planck distribution \cite{CMBR}.
The corresponding constraint is, however, rather milder than that from the abundance of $^3$He/D. 
\item $10^{12} \sec \lesssim \tau_{\phi_b} \lesssim 10^{24} \sec$:
In this case, moduli decay after the recombination era.
The most stringent constraint is coming from the diffused X($\gamma$)-ray 
\cite{Choi:1998dw,Churazov:2006bk}.
For the life-time longer than the age of the universe ($\tau_{\phi_b} \gtrsim 10^{17} \sec$), 
X($\gamma$)-ray produced in the galactic center should be also considered.
\item $\tau_{\phi_b} \gtrsim 10^{24} \sec$:
In this case, the only constraint is from the present dark matter relic density, $\Omega_{\rm DM}h^2 \simeq 0.11$ \cite{PDG}. 
\end{itemize}
All these constraints are shown in Fig.~\ref{fig:mod_problem} where we see that, for $10 \MeV \lesssim m_{\phi_b} \lesssim 100 \TeV$, the abundance of the light modulus is constrained to be less than 
\beq
\Omega_{\phi_b}h^2 \lesssim \mathcal{O}(10^{-8} - 10^{-2})
\eeq
that corresponds to 
\beq \label{Y-constraint}
\frac{n_{\phi_b}}{s} \lesssim \mathcal{O}(10^{-16} - 10^{-10}) 
\eeq
where $n_{\phi_b}$ and $s$ are the number density of the light modulus and entropy density, respectively.
%

The cosmological moduli problem is that the typically expected moduli abundance 
is much larger than the upper bounds shown in Fig.~\ref{fig:mod_problem}.
In the very early universe, when the expansion rate becomes comparable to the mass of a modulus, the modulus is expected to be produced enormously in the form of coherent oscillation caused by vacuum misalignment.
In case of the small volume modulus, the initial oscillation amplitude is expected to be string scale, and the modulus never dominates the universe until it decays well before BBN.
Hence it is harmless.
On the other hand, the large volume modulus has Planckian initial oscillation amplitude
\footnote{The initial misalignment of $\phi_b$ might be typically super-Planckian.
For such a large field value, the potential is of an exponential form.
Then, in the universe dominated by radiation it will be quickly attracted to near its vacuum position \cite{Conlon:2008cj}.
Once $\Delta \phi_b \lesssim M_{\planck}$ with respect to the vacuum position, the potential can be approximated as a quadratic form, and the modulus oscillates with an amplitude of order Planck scale and behaves like a matter.
Therefore, we can simply assume that the initial misalignment of the light modulus is of order Planck scale when it starts to oscillate in a radiation dominated universe.
}
, and dominates the universe right after it starts to oscillate or at least when it decays.

We assume that the expansion rate during the primordial inflation is much larger than the mass of the small modulus.
The light modulus starts to oscillate when $H \sim m_{\phi_b}$. 
Denoted as $Y \equiv n_{\phi_b} / s$, the late-time abundance of the light modulus is given by
\beq \label{BB-moduli}
Y_{\rm BB,0} = \frac{3}{4} \times \left\{
\begin{array}{lcc}
\left[ \frac{\pi^2}{30} g_*(T_*) \right]^{-1/4} \left( \frac{M_\planck}{m_{\phi_b}} \right)^{1/2} & \textrm{for} & \Gamma_I \gtrsim m_{\phi_b}
\\
\frac{T_{\rm R}}{m_{\phi_b}} & \textrm{for} & \Gamma_I < m_{\phi_b}
\end{array}
\right.
\eeq
where $T_{\rm R}$ is the reheating temperature of inflaton and defined as
\beq
T_{\rm R} \equiv \left[ \frac{\pi^2}{90 g_*(T_{\rm R})} \right]^{-1} \left( \Gamma_I M_\planck \right)^{1/2}
\eeq
with $\Gamma_I$ being its decay rate, and we assumed that the energy density of the modulus is equal to that of inflaton for $\Gamma_I \lesssim m_{\phi_b}$.
If $\Gamma_I \gtrsim m_{\phi_b}$, the late-time entropy release in the decay of heavy moduli, $\phi_s$, can dilute the abundance of the light modulus by a factor
\beq
\Delta_s \simeq \left( \frac{\gamma_{s \to A_\mu A_\mu}}{64 \pi} \right)^{-1} \left( \frac{m_{\phi_b}}{m_{\phi_s}} \right)^{5/8} \left( \frac{M_{\rm st}}{M_\planck} \right)^{5/2} \left( \frac{M_\planck}{m_{\phi_b}} \right)
\eeq
at most.
Hence the late-time abundance of the light modulus after the decay of the small volume modulus is 
\beq \label{BB-mod-HighTR}
Y_{\rm BB} = Y_{\rm BB, 0} \frac{1}{\Delta} 
\simeq 0.1 \gamma_{s \to A_\mu A_\mu} \left( \frac{m_{\phi_b}}{1 \GeV} \right)^{1/2}
\eeq
where we used $\tau_b\simeq 10^4$ and $\ln(M_\planck / m_{3/2})\simeq 4\pi^2$.
If $\Gamma_I < m_{\phi_b}$, we find 
\beq \label{BB-mod-LowTR}
Y_{\rm BB} \gtrsim 10^{-3} \left( \frac{T_{\rm R}}{1 \MeV} \right) \left( \frac{1 \GeV}{m_{\phi_b}} \right)
\eeq
since the reheating temperature of inflation should be larger than about $1 \MeV$ for successful BBN.
As shown in \eqs{BB-mod-HighTR}{BB-mod-LowTR},  irrespective of the reheating temperature of inflation, the abundance of the large volume modulus is too much to match the observational constraints, \eq{Y-constraint}.
Hence, the light modulus causes a disaster in a consistent cosmology unless its abundance is somehow diluted enough.

\section{Thermal inflation}
The most compelling solution to the cosmological moduli problem is thermal inflation \cite{Lyth:1995hj}.
In this section, we show that single thermal inflation is only a partial solution to the moduli problem in LVS for the mass scale of moduli(\eq{m-phib}), though it depends on the reheating temperature of the primordial inflation.
We then consider a double thermal inflation as a complete solution to the problem, and clarify how it works.

\subsection{A single thermal inflation}
The flaton field, denoted as $X$, controlling thermal inflation has a zero-temperature potential,
\beq
V(X) = V_0 - m_X^2 |X|^2 + \cdots
\eeq
where ($\cdots$) represent possible higher order term(s) to stabilize $X$.\footnote{
In the LVS, it is difficult to obtain such a  flaton field ($X$) at low energy in the closed 
string moduli sector \cite{Anguelova:2009ht}. 
Instead, $X$ is considered as a open string mode along with the SM matter fields. }  
Then, denoting the VEV of $X$ as $X_0$ and requiring vanishing cosmological constant at true vacuum, one finds
\bea \label{STI}
V_0 &\sim& m_X^2 X_0^2
\\
\Gamma_X &=& \frac{1  }{8 \pi}\gamma_X \frac{m_X^3}{X_0^2},\quad
T_{\rm d} \equiv \left( \frac{\pi^2}{90} g_*(T_{\rm d}) \right)^{-1/4} \left( \Gamma_X M_\planck \right)^{1/2}
\eea
where $m_X$ is the physical mass of the flaton, $\Gamma_X$ is the decay rate of $X$, 
 $\gamma_X$ is a factor determined by the coupling of $X$ to SM particles and not a function of $X_0$, and $T_{\rm d}$ is the decay temperature of $X$, which is assumed to be dominated by SM particles to re-estabilish radiation background for a successful BBN.
Thermal inflation takes place when $V_0$ dominates the energy density of the universe while the background temperature is still larger than the critical temperature $T_{\rm c} \sim m_X$ at which $X$ is destabilized from the origin and ends thermal inflation. 

The epoch of flaton domination follows thermal inflation, and the eventual decay of flaton reheats the universe, releasing huge amount of entropy.
As the result, moduli are diluted by a factor 
\beq \label{TI-dilution}
\Delta
\simeq \frac{g_{*S}(T_{\rm d})}{g_{*S}(T_{\rm c})} \left[ \frac{\pi^2}{30} g_*(T_{\rm d}) \right]^{-1} \frac{V_0}{T_{\rm c}^3 T_{\rm d}}
\eeq
However there is some amount of moduli reproduction in the following way.
During thermal inflation, the moduli potential is of the form 
\bea
V({\phi_b}) 
&=&\frac{1}{2} m_{\phi_b}^2 {\phi_b}^2 + \frac{c_b{\phi_b}}{M_\planck} V_0 + \dots
\\
&=& \frac{1}{2}m_{\phi_b}^2 \left({\phi_b}+  \frac{c_bV_0}{m_{\phi_b}^2 M_\planck} \right)^2 + \dots
\eea
where $c_b$ is assumed as a constant of order unity ($c_b=1/3$ has been used throughout this paper). 
Hence moduli is shifted by the amount of 
\beq
\delta {\phi_b}\sim \frac{ c_b  V_0}{m_{\phi_b}^2 M_\planck} 
\eeq
and reproduced after thermal inflation with the amount of  
\beq \label{TI-moduli}
Y_{\rm TI, 0} 
= \frac{1}{2}\left( \frac{2 \pi^2}{45} g_{*s}(T_{\rm c}) \right)^{-1} \frac{m_{\phi_b}\delta \phi_b^2}{T_{\rm c}^3}
\eeq
Therefore, the late-time abundance of moduli is the sum of \eqs{BB-moduli}{TI-moduli}, that is 
\footnote{When thermal inflation begins, the contribution of the small volume modulus to entropy is negligible.}
\beq
Y_{\rm tot} = \left[ Y_{\rm BB,0} + Y_{\rm TI,0} \right] \frac{1}{\Delta}
\eeq
In the right-hand side of the above equation, the first and second terms as functions of $X_0$ provide decreasing and increasing contributions, respectively.
Hence the total moduli abundance can have a minimum at $X_0=X_0^{\rm min}$.  
Depending on the decay rate of inflaton, it is minimized as follows.
\paragraph{$\Gamma_I \gtrsim m_{\phi_b}$:}
The abundance of moduli is minimized at $X_0^{\rm min}$ satisfying
\beq
Y_{\rm TI,0} = 3 Y_{\rm BB,0}
\eeq
with the minimum abundance,
\beq \label{Yb-in-STI-HighTR}
Y_{\rm tot}^{\rm min} = 4 Y_{\rm BB,0} \frac{1}{\Delta}.
\eeq
Since $Y_{\rm BB} \propto m_{\phi_b}^{-1/2}$ and $X_0^{\rm min} \propto m_{\phi_b}^{5/8}$, we see that the minimum abundance of the light moduli increases as $m_{\phi_b}$ decreases.
\paragraph{$\Gamma_I < m_{\phi_b}$:}
In this case, thermal inflation can take place if $T>T_{\rm c}$ when the energy density of inflaton is comparable to $V_0$.
Then, as long as the energy density of inflaton is subdominant at its decay (i.e., $\Gamma_X < \Gamma_I$), $Y_{\rm BB, 0}$ is minimized for $T_{\rm R} \sim V_0^{1/4}$.
That is,
\beq \label{YBB0min}
Y_{\rm BB, 0} \gtrsim \frac{V_0^{1/4}}{m_{\phi_b}}
\eeq
Then, taking the minimum value of \eq{YBB0min}, we find
\beq 
5\, Y_{\rm BB,0} = 2\, Y_{\rm TI,0}
\eeq
at $\ X_0=X_0^{\rm min}$ with 
\beq \label{Yb-in-STI-LowTR}
Y_{\rm tot}^{\rm min} \gtrsim \frac{7}{2} Y_{\rm BB,0} \frac{1}{\Delta}.
\eeq
The abundance of moduli increases as $m_{\phi_b}$ decreases in this case too.

\begin{figure}[ht] 
\centering
\includegraphics[width=0.45\textwidth]{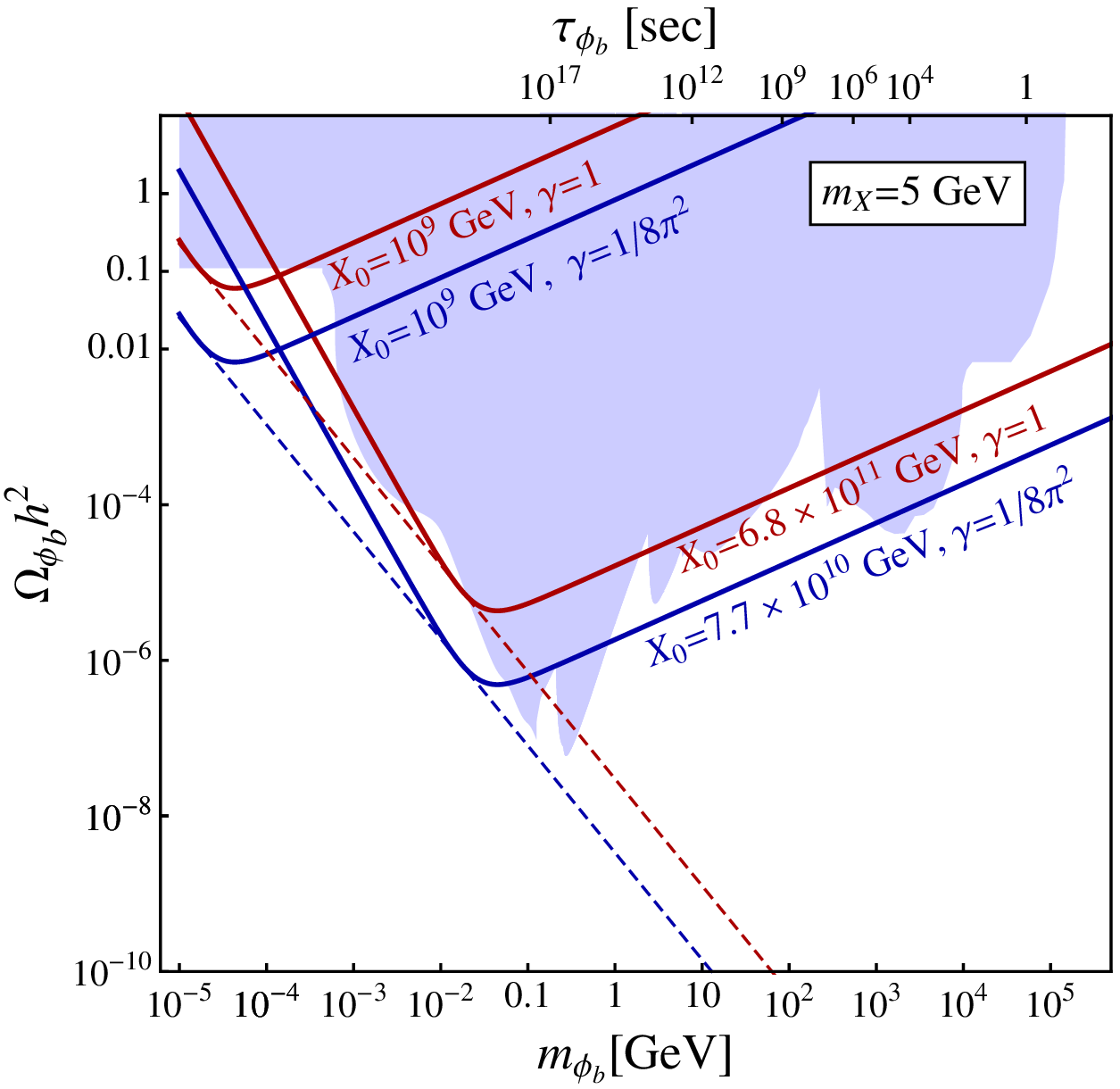} 
\includegraphics[width=0.45\textwidth]{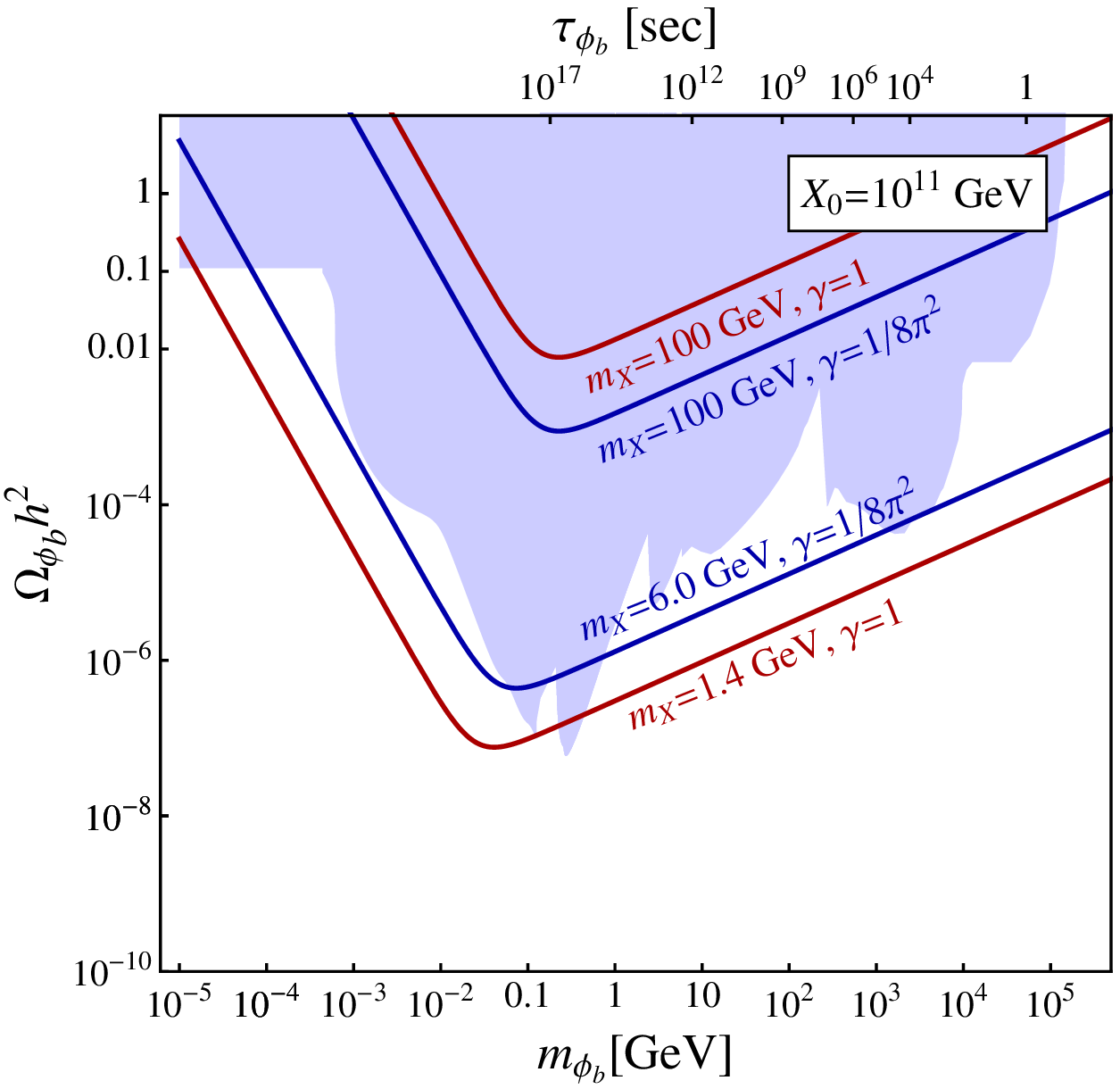} 
\includegraphics[width=0.45\textwidth]{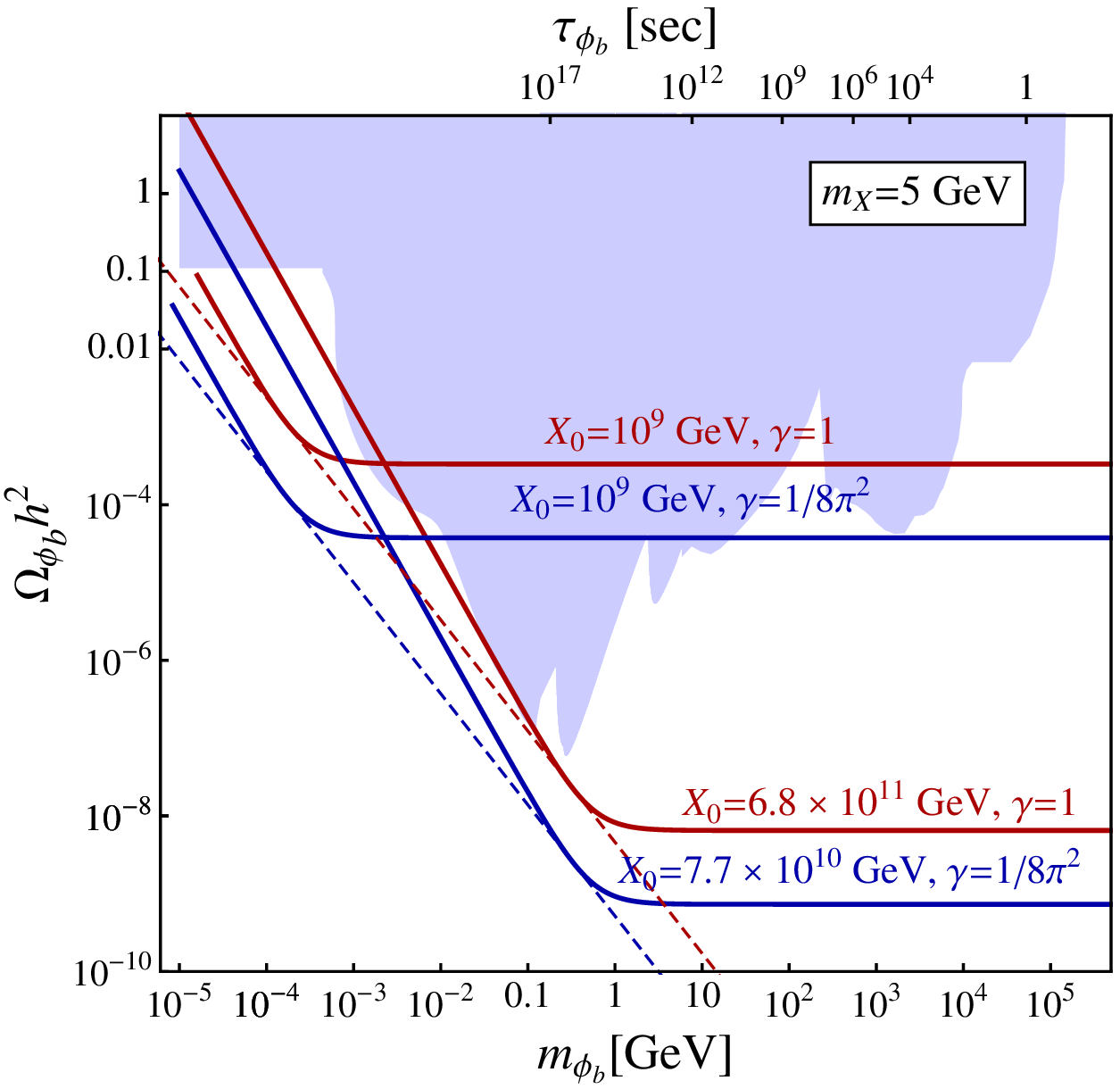} 
\includegraphics[width=0.45\textwidth]{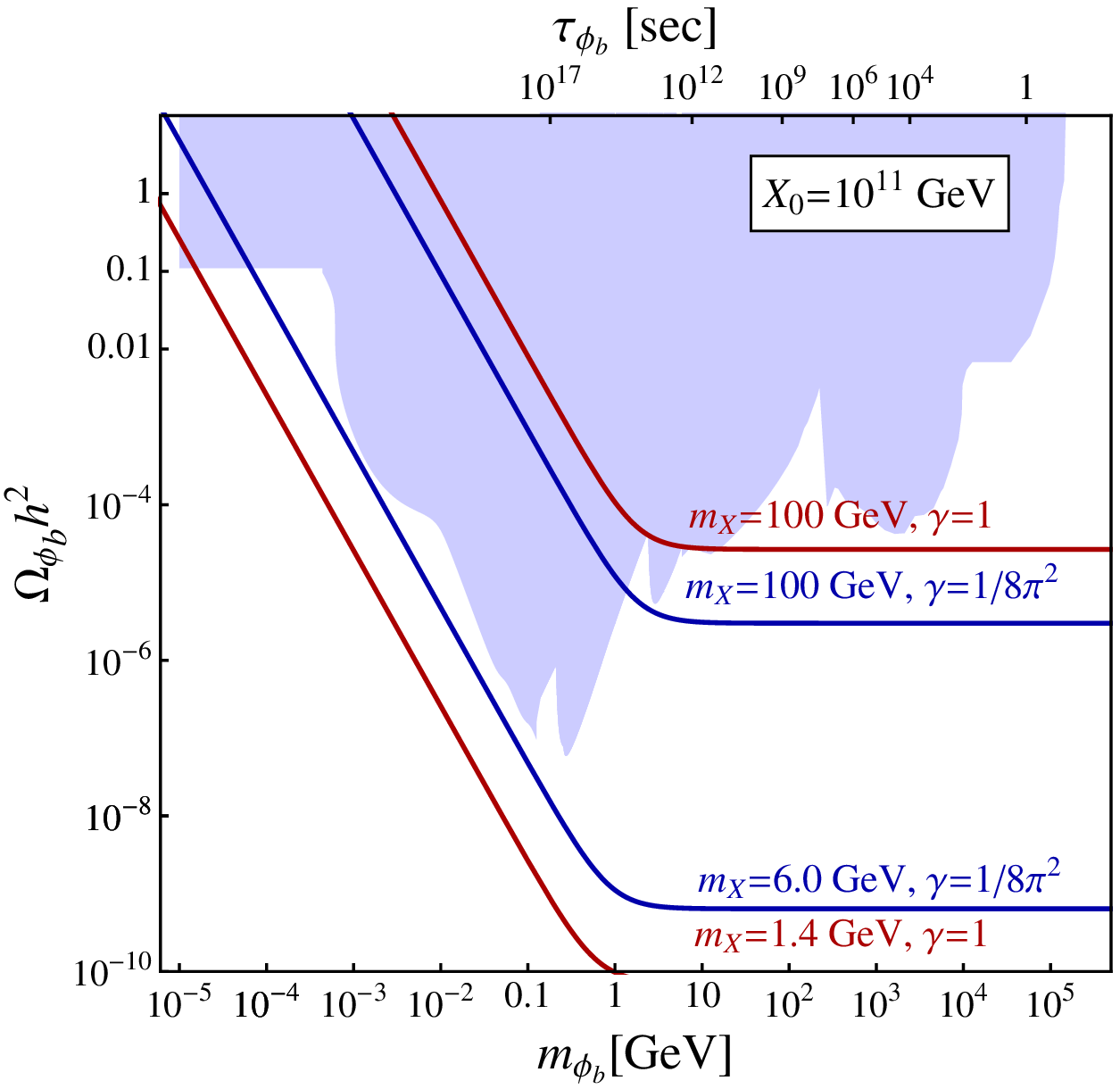} 
\caption{Top: $\Gamma_I \gtrsim m_{\phi_b}$. Bottom: $\Gamma_I = V_0^{1/2} / \sqrt{3} M_\planck$ maximizing dilution effect from inflation. Left: a fixed flaton mass, $m_X = 5 \GeV$. Right: a fixed flaton VEV, $X_0 = 10^{11} \GeV$. 
The light blue colored region is excluded by observations, and the moduli relic abundance $\Omega_{\phi_b}h^2$ after a single thermal inflation was depicted as various (dotted, dashed, solid) lines depending on the choice of parameter set. 
Dark red and blue lines are respectively the case of $\gamma_X=1$ and $\gamma_X=1/8\pi^2$ in 
Eq.~(\ref{STI}).  Solid lines correspond to the case of minimal decay temperature $T_{\rm d}=5\MeV$.
Dotted lines of the left panels correspond to the minimum abundances of $\phi_b$ given by \eq{Yb-in-STI-HighTR} and~(\ref{Yb-in-STI-LowTR}), respectively. }
\label{fig:moduli-in-STI}
\end{figure}

In \eqs{Yb-in-STI-HighTR}{Yb-in-STI-LowTR}, one crucial factor affecting the abundance of moduli is the decay rate which depends on the coupling of $X$ to SM particles.
For an intermediate scale $X_0$, there are two leading possibilities.
\begin{itemize}
\item $\mu$-term coupling:
\beq
W \supset \frac{\lambda_\mu }{nM^{n-1}}\,  X^nH_u H_d
\eeq
The energy density of the flaton $X$ soon after thermal inflation is nearly equally distributed to the radial and axial modes.
If there is a symmetry under which $X$ is charge, and the symmetry is broken only spontaneously, then only the radial mode is relevant in our argument (though the axial mode is likely to be the axion for strong-CP problem of QCD).
The decay rate of $|X|$ in this case is 
\beq
\Gamma_{|X|} \sim \frac{3}{4 \pi} \frac{|\mu|^4}{m_X X_0^2}
\eeq
It gives too high decay temperature to solve the problem of the light moduli.

If $X$ is stabilized by higher order term(s), for example, a self-coupling, then radial and axial modes have masses similar to each other.
For  $m_X \sim m_{a_X} \lesssim m_{\rm soft}$ with $m_{a_X}$ being the mass of axial component of $X$, the axial mode decays later than radial mode.
The decay rate is given by
\beq \label{mu-term-decay}
\Gamma_a \simeq \sum_i \frac{N_{{\rm c}, i}}{8 \pi} \frac{m_{a_X}^3}{X_0^2} \left( \frac{|B \mu|^2}{m_A^4} \right) \left( \frac{m_{d_i}^2 \tan^2 \beta}{m_{a_X}^2} \right) \left( 1 - \frac{4 m_{d_i}^2}{m_{a_X}^2} \right)^{3/2}
\eeq
where $N_{{\rm c}, i}$ is the color factor for a particle $i$, $B$ is the soft parameter of the Higgs bilinear term, $m_A$ the mass of the CP-odd Higgs and $m_{d_i}$ is the mass of a particle coupled to down-type Higgs.
\item Hadronic coupling:
\beq
W \supset \lambda X \Psi \bar{\Psi}
\eeq
where $\Psi$ and $\bar{\Psi}$ are assumed SM charged fields.
The effective coupling of the radial and axial modes of $X$ to SM sector is 
\beq
\lambda_{\rm eff} \sim \frac{g^2}{16 \pi^2} \frac{1}{X_0}
\eeq
where $g$ is the gauge coupling of $\Psi$ and $\bar{\Psi}$.
If there is a global symmetry under which $X$ is charged, there will be 
very light Goldstone boson to which $|X|$ decays dominantly.
This is dangerous because BBN requires a universe dominated by SM-like radiation.
Therefore, global symmetries should be badly broken. 
In this case, radial and axial mode have mass scales similar to each other, and the decay rate is given by
\beq \label{decay-via-hadron}
\Gamma = \frac{1}{8 \pi} N_\Psi \left( \frac{g^2}{16 \pi^2} \right)^2 \frac{m_X^3}{X_0^2}
\eeq
where $N_\Psi$ is the number of the pairs of $\Psi + \bar{\Psi}$.
The perturbativity of gauge coupling restricts $N_\Psi$ to be 
less than about 8 if $\Psi$ ($\bar{\Psi}$) is a fundamental 
representation of $SU(5)$ \cite{Morrissey:2005uz}. 
\end{itemize} 
Using \eqs{mu-term-decay}{decay-via-hadron}, we show the effect of single thermal inflation on the moduli problem in Fig.~\ref{fig:moduli-in-STI}.
%
In the figure, the red and blue lines corresponds to flaton decay via $\mu$-term and hadronic interactions, respectively.
The solid lines are for specific choices of parameters shown in the figure, while dashed lines are for minimum abundances of moduli obtained by using $X_0^{\rm min}$ for given sets of mass parameters.   
It is clear that the case of hadronic interaction is more efficient in solving moduli problem.
Such an efficiency is due to the low decay temperature originated from the smallness of $\gamma_X$.
In the upper panels of the figure, where we assumed $\Gamma_I \gtrsim m_{\phi_b}$, we notice that if $m_X \gtrsim 100 \GeV$ only $m_{\phi_b} \gtrsim 1 \GeV$ can be viable whereas, for $m_X \lesssim 5 \GeV$, $m_{\phi_b} \lesssim 10^{-2}$ can also be saved.
Saving $m_{\phi_b} = \mathcal{O}(10^{-2} - 1) \GeV$ is possible only if $T_{\rm d}$ is pushed down to a few $\MeV$.
Note that for such a low decay temperature of flaton there may be no way to generate right amount of baryon number asymmetry though dark matter may be obtained from axions with a large coupling constant.
However, as shown in the lower panel, if $\Gamma_I$ is close to $V_0^{1/2}/ \sqrt{3} M_\planck$ for which inflaton decay has maximal effect on the abundance of moduli, single thermal inflation can solve the moduli problem for the whole range of $m_{\phi_b}$ much easily.
In addition, there may be a room for baryogenesis, such as late-time Affleck-Dine leptogenesis. 

Consequently, we notice that if $\Gamma_I \sim V_0^{1/2} / \sqrt{3} M_\planck$ the moduli problem of the light moduli in LVS can be solved by a single thermal inflation for the whole relevant mass range.
However such a possibility is the most optimistic case and might not be realized in nature.
Hence, generically, a single thermal inflation can solve the moduli problem of the light moduli in LVS only partially.
Particularly, the mass scale of moduli associated with $\TeV$ scale gravitino mass is not viable in general.

\subsection{Double thermal inflation}
\label{subsec:DTI}
In this section, as a minimal extension of a single thermal inflation to solve the moduli problem completely, we consider a double-stage thermal inflation and clarify how it works.
%
Here, we assume $\Gamma_I \gtrsim m_{\phi_b}$ for simplicity
\footnote{In the case of $\Gamma_I < m_{\phi_b}$, the minimal abundance of moduli is given in the same way as the case of the single thermal inflation.}.

Compared to a single-stage thermal inflation, the existence of the second stage of thermal inflation is useful for diluting moduli produced before the second thermal inflation.
Hence, as long as we can manage for the second thermal inflation not to reproduce too much moduli, the double thermal inflation would be able to solve the cosmological problem of the light moduli.
Since the extra dilution we need from the second thermal inflation may not be very big, a bit high reheating temperature after the second thermal inflaiton might be possible, and this may open a room for late-time Affleck-Dine mechanism to work.

Let us denote the two flat directions which trigger the end of each thermal inflation as $X_1$ and $X_2$, and the potential energy of flatons as
\beq
V(X_1,X_2) = \left\{
\begin{array}{ccl}
V_1 & {\rm for} & \langle X_1 \rangle = \langle X_2 \rangle = 0
\\
V_2 & {\rm for} & \langle X_1 \rangle = X_{1,0}, \ \langle X_2 \rangle = 0
\end{array}
\right.
\eeq
where $\langle \dots \rangle$ is the vacuum expectation value (VEV) and $X_{1,0}$ is the VEV of $X_1$ at zero temperature true vacuum.
We assume 
\beq \label{DTI-cond0}
V_1 \gg V_2
\eeq
Then, $V_1$ and $V_2$ may drive the first and second thermal inflation, respectively.
In order to have such an order of thermal inflation, we require  
\beq \label{DTI-cond1}
T_{\rm c, 1} > T_{\rm c, 2}
\eeq
where $T_{\rm c, i}$ is the critical temperature when $X_i$ is destabilized from the origin.
If this condition is not satisfied, $X_2$ would be destabilized at or before the end of the first thermal inflation and the second thermal inflation will not take place.
Since $T_{\rm c, i} \sim m_{X_i}$ for order unity coupling to thermal bath with $m_{X_i}$ being the tachyonic curvature of the zero temperature potential along $X_i$ around the origin, \eq{DTI-cond1} implies 
\beq
m_{X_1} > m_{X_2}
\eeq
In addition to the condition \eq{DTI-cond1}, the background temperature when the second thermal inflation begins should be larger than $T_{\rm c, 2}$, otherwise $X_2$ would be destabilized before $V_2$ starts to dominate and the second thermal inflation would not take place.
Hence another condition for the second thermal inflation is  
\beq \label{DTI-cond2}
T(t_2) > T_{\rm c, 2}
\eeq
where $T(t_2)$ is the temperature of the radiation which couples to $X_2$ at the epoch when the second thermal inflation begins.
The radiation density after the first thermal inflation is given by
\beq
\rho_{\rm r} = \rho_{\rm r, bg} + \Delta \rho_{\rm r}
\eeq
where $\rho_{\rm r, bg}$ is the background radiation and $\Delta \rho_{\rm r}$ is the radiation contribution from the partial decay of $X_1$.
Soon after the end of the first thermal inflation, one finds
\beq
\Delta \rho_{\rm r} \simeq \frac{2}{5} \frac{\Gamma_{X_1 \to {\rm SM}}}{H} \rho_1
\eeq
where $\Gamma_{X_1 \to {\rm SM}}$ is the partial decay width of $X_1$ to SM particles which couple efficiently to $X_2$ directly or indirectly, and $\Gamma_{X_1 \to {\rm SM}} \ll H$ was assumed. 
This should dominate over $\rho_{\rm r, bg}$ at least before the time $t_2$.
Therefore, we find 
\beq
\rho_{\rm r}(t_2) 
\sim \frac{\Gamma_{X_1 \to {\rm SM}}}{H_2} V_2
= {\rm Br}(X_1 \to {\rm SM}) \left( \frac{\pi^2}{30} g_*(T_{\rm d, 1}) \right)^{1/2} T_{\rm d, 1}^2 V_2^{1/2}
\eeq
where $T_{\rm d, 1}$ is defined as
\beq
\frac{\pi^2}{30} g_*(T_{\rm d, 1}) T_{\rm d, 1}^4 = 3 \Gamma_1^2 M_\planck^2
\eeq
with $\Gamma_1$ being the total decay rate of $X_1$.
Hence \eq{DTI-cond2} is translated to 
\beq
\Gamma_{X_1 \to {\rm SM}} > \frac{1}{\sqrt{3}} \frac{\pi^2}{30} g_*(T(t_2)) \frac{T_{\rm c, 2}^4}{V_2^{1/2} M_\planck}
\eeq
For intermediate scale VEVs of $X_1$ and $X_2$, and $m_{X_1} \gtrsim m_{X_2} \sim T_{\rm c, 2}$, this condition is easily satisfied.
Therefore, double thermal inflation can be realized easily.

The number of $e$-foldings is given by
\beq
N_{e, \rm tot} = \ln \left( \frac{T(t_1)}{T_{\rm c, 1}} \right) + \frac{4}{3} \ln \left( \frac{T(t_2)}{T_{\rm c, 2}} \right)
\eeq
where $T(t_1)$ is the temperature when the first thermal inflation begins, and the first/second term in the right-hand side is the contribution of the first/second thermal inflation.
The dilution factors due to entropy release of both stages of thermal inflation are given by 
\bea
\Delta_1
&\simeq& \frac{g_{*s}(T(t_2))}{g_{*s}(T_{\rm c, 1})} \frac{T(t_2)^3}{T_{\rm c,1}^3} \left( \frac{a(t_2)}{a_{\rm c ,1}} \right)^3
\simeq \mathbb{C}_1 {\rm Br}(X_1 \to XX)^{3/4} \frac{T_{\rm d, 1}^{3/2}}{T_{\rm c, 1}^3} \frac{V_1}{V_2^{5/8}}
\label{TI1-dilution} \\
\Delta_2 
&\simeq& \mathbb{C}_2 \frac{V_2}{T_{\rm c, 2}^3 T_{\rm d, 2}} 
\label{TI2-dilution}
\eea
where 
\bea
\mathbb{C}_1 &\equiv& \frac{g_{*s}(T(t_2))}{g_{*s}(T_{\rm c, 1})} \left( \frac{\pi^2}{30} g_*(T(t_2)) \right)^{-3/4} \left( \frac{\pi^2}{30} g_*(T_{\rm d, 1}) \right)^{3/8} 
\\
\mathbb{C}_2 &\equiv& \frac{g_{*S}(T_{\rm d, 2})}{g_{*S}(T_{\rm c, 2})} \left[ \frac{\pi^2}{30} g_*(T_{\rm d, 2}) \right]^{-1}
\eea
%
The late-time total abundance of moduli is 
\beq
Y_{\rm tot} = \left[ \left( Y_{\rm BB,0} + Y_{\rm TI1,0} \right) \frac{1}{\Delta_1} + Y_{\rm TI2,0} \right] \frac{1}{\Delta_2}
\eeq
where $Y_{\rm BB,0}$ is given by \eq{BB-moduli}, and 
\beq
Y_{\rm TIi} = \frac{1}{2} \left( \frac{2 \pi^2}{45} g_{*s}(T_{\rm c}) \right)^{-1} \frac{c_b^2V_i^2}{T_{\rm c,i}^3 m_{\phi_b}^3 M_\planck^2}
\eeq
with $i=1,2$.
It is minimized at $X_{1,0}^{\rm min}$, $X_{2,0}^{\rm min}$ satisfying
\beq
Y_{\rm TI1,0} = \frac{1}{7} Y_{\rm BB,0}\, , \quad Y_{\rm TI2,0} =  \frac{2}{\Delta_1}\, Y_{\rm BB,0}
\eeq
with the minimum abundance given by 
\beq \label{YtotMin-in-DTI}
Y_{\rm tot}^{\rm min} = \frac{22}{7} \frac{Y_{\rm BB,0}}{\Delta_1 \Delta_2}.
\eeq

If the decay temperature of the second thermal inflation is high enough, there is a chance for the late-time Affleck-Dine leptogenesis to provide a right amount of baryon number asymmetry.
In order for the mechanism to work, the flaton which triggers the end of the last thermal inflation should trigger the mechanism too, otherwise it is difficult to obtain a dynamics triggering the mechanism.
This implies that Affleck-Dine field should be in a symmetry-breaking phase when the last thermal inflation ends, that is 
\beq
T_{\rm c, AD} > T_{\rm c, 2}
\eeq
and $X_2$ should be able to lift up the AD field. 
We may also have a good dark matter candidate, for example the fermion superpartner of $X_2$, although QCD-axion might be still a good candidate.   

As a brief summary of this section, a double thermal inflation is possible if following inequalities are satisfied. 
\beq
V_1 \gg V_2 \quad , \quad T_{\rm c, 1} > T_{\rm c, 2} \quad , \quad T(t_2) > T_{\rm c, 2}
\eeq

\section{A model}

We now consider a concrete example which can realize a double thermal inflation.
Our model is characterized by the following superpotential.
\bea \label{W-DTI}
W 
&=& W_{\rm MSSM} + \frac{\lambda_\mu}{M_*} Z^2 H_u H_d + \frac{1}{2} \frac{\lambda_\nu}{M_\nu} \left( L H_u \right)^2 + \frac{\lambda_Z}{4 M_*} Z^4 + \frac{1}{2} \lambda_S Z S^2 
\nonumber \\
&& \phantom{W_{\rm MSSM}} + \lambda_\Psi X \Psi \bar{\Psi} + \lambda_\Phi Y \Phi \bar{\Phi} + \frac{\kappa}{3 M_*} X^3 Y
\eea
where $W_{\rm MSSM}$ is the MSSM superpotential without $\mu$-term, 
$\Psi$ ($\bar{\Psi}$) and $\Phi$ ($\bar{\Phi}$) are SM charged matter fields, 
$X$, $Y$, $Z$ and $S$ are gauge-singlets under the SM gauge groups. 
An anomalous global $U(1)_{\rm PQ}$ symmetry is introduced  to provide a flaton field ($X$) responsible for the first thermal inflation and solve the strong CP problem \cite{Kim:1979if} simultaneously.
$Z$ is introduced  to  trigger the second thermal inflation.
It reproduces the Higgs bilinear $\mu$-term by its intermediate scale VEV. 
We impose an appropriate discrete $\mathbb{Z}_4$ symmetry as  in Table \ref{tab:Z4charge}.
The coupling of the seesaw operator is constrained so that the mass of left-handed neutrino is given by
\beq
m_\nu = \frac{\lambda_\nu v_u^2}{M_\nu}
\eeq
where $v_u$ is the VEV of up-type neutral Higgs field.
All $\lambda_i$s are assumed to be of order unity with $M_* \sim M_{\rm GUT}$. 
%

\begin{table}[htdp]
\caption{charges under the $U(1)_{\rm PQ}$ and discrete $\mathbb{Z}_4$ symmetry}
\begin{center}
\begin{tabular}{|c||c|c|c|c|c|c|c|c|}
\hline
Field & $H_uH_d$ & $LH_u$  &$ \Psi\bar\Psi $ & $\Phi\bar\Phi$ &$X $ &$Y$ & $Z$ & $S^2$  
\\ \hline
$U(1)_{\rm PQ}$& $0$ & $0$  & $1$ & $-3$ 
 & $-1$ & $3$   & $0$ & $0$ \\ \hline
$\mathbb{Z}_4$ & 2 & 2  & 0 & 0  & 0 & 0 & 1 & 3 
\\ \hline
\end{tabular}
\end{center}
\label{tab:Z4charge}
\end{table}
Since $Z$ has a non-zero vacuum value, domain walls due to the spontaneously broken discrete $\mathbb{Z}_4$ symmetry appears and becomes problematic if they are stable.
However the symmetry might not be exact due to either anomaly or higher order symmetry-breaking terms originated from gravitational violation of global symmetry.
Hence the walls are unstable unless the $\mathbb{Z}_4$ is a discrete gauge symmetry without anomaly.
Collapsing walls produces a gravitational wave background, whose energy density is constrained by observations.
For an intermediate scale VEV of $Z$, observational constraints requires a symmetry-breaking stronger than anomaly \cite{Moroi:2011be}.

\subsection{Potential of singlet scalars and gauge mediation}
We assume that $S$ couples somehow to SM-charged non-SM field(s) so that it is never destabilized.
Ignoring the VEV of Higgs fields, the potential of the singlets is given by 
\beq
V = V_0 + V_{\rm PQ}(X,Y) + V_Z(Z)
\eeq
where $V_0$ is for present vanishing cosmological constant, and $V_{\rm PQ}$ and $V_Z$ are the scalar potential of the {\it canonically normalized} PQ breaking fields and $Z$ field, respectively.
We find 
\bea
V_{\rm PQ} &=&  m_{X}^2|X|^2 + m_Y^2|Y|^2 + \left(\frac{A_\kappa \kappa}{3 M_{\rm  GUT}} X^3Y + {\rm
h.c}\right)
\nonumber \\
&& + \frac{|\kappa|^2}{M_{\rm GUT}^{2}}|X|^{6} + \frac{|\kappa|^2}{M_{\rm GUT}^{2}}|X|^{4}|Y|^2
\\
V_Z &=&  m_Z^2 |Z|^2 + \left( \frac{A_Z \lambda_Z}{4 M_{\rm GUT}} Z^4 + \mathrm{c.c.} \right) + \frac{|\lambda_Z|^2}{M_{\rm GUT}^2} |Z|^6
\eea
where $M_{\rm GUT}$ is the apparent GUT scale for the visible sector in LVS. 
Soft SUSY breaking terms ($m_{X,Y,Z}^2$, $A_{\kappa}$, and $A_Z$)
are quiet model dependent.  
We consider a natural set-up realized in string theory \cite{Blumenhagen:2007sm}, 
where the visible sector volume modulus is stabilized by 
$D$-term potential from the pseudo anomalous $U(1)_A$ gauge symmetry. 
In this case,  the $U(1)_{\rm PQ}$  can be a part of  $U(1)_A$
and the corresponding SUSY breaking $D$-term ($D_A$) 
is obtained as of the order of gravitino mass  \cite{Choi:2010gm}.
Then,  $D$-term mediated soft scalar masses are proportional to 
the $U(1)_A$ charge as $m_{\varphi_i}^2 = q^A_i D_A$, 
where $q^A_i $ is the same as the charge under the $U(1)_{\rm PQ}$, A-parameters and gaugino masses are generically suppressed with respect to $D_A$,
so that $|A_\kappa|, |A_Z|\,\ll\,|m_X|, |m_Y|$.  
In this set-up, $U(1)_A$ charges are imposed as  $m_Y^2= - 3 m_X^2 \, > \, 0$.
We then find 
\bea |X| &\simeq&
\left(\frac{\sqrt{3} |m_X|M_{\rm GUT}}{|\kappa|}\right)^{1/2},\quad
\left|\frac{Y}{X}\right| \,\simeq \,
\frac{1}{6 \sqrt{3}}\left|\frac{A_\kappa}{m_X}\right|,
\eea 
and also the SUSY breaking auxiliary components of $X$ and $Y$ are
\bea
\left|\frac{F^X}{X}\right|
\,\simeq\, \frac{1}{6}|A_\kappa|,\quad 
\left|\frac{F^Y}{Y}\right| &\simeq& 6
\left|\frac{m_X^2}{A_\kappa}\right|.
\eea
Here a particulary interesting feature is that $F^Y/Y$ is enhanced by $m_X/A_\kappa$ in the limit $|A_\kappa|\ll |m_X|$. 
In LVS with $U(1)_A$,  we  have 
\bea
|m_X| \,\simeq\, \sqrt{D_A} \,=\, {\cal O}\left(8\pi^2 A_\kappa\right), 
\eea 
for which  
\bea
\frac{1}{8\pi^2} \frac{F^Y}{Y}\,\sim\, 6\sqrt{D_A}={\cal O}(m_{3/2}).
\eea 
The presence of the superpotential tem
\beq
W \supset \lambda_\Phi Y \Phi \bar{\Phi}
\eeq
allows a gauge mediation of the SUSY breaking by $F^Y$.
If there are  $N_\Phi$ flavors of $\Phi+\bar{\Phi}$ which form the $5+\bar 5$ of $SU(5)$, the gauge threshold contributions just below  the messenger scale $M_{\rm mess}=|\lambda_\Phi Y| $ to the soft parameters are given by
\bea\label{gauge}
\Delta m_{\varphi_i}^2 = 2N_\Phi {\rm Tr}(T_a^2(\Phi_i))
\left|\frac{g_a^2}{16\pi^2}\frac{F^Y}{Y}\right|^2,\quad  
  M_a = -N_\Phi \left(\frac{g_a^2}{16\pi^2}\frac{F^Y}{Y}\right).
\eea
Then the set-up gives  
\bea 
v_{\rm PQ}&\sim& |X| \,\sim\,
\left(\frac{3 \sqrt{D_A} M_{\rm GUT}}{\kappa}\right)^{1/2},
\quad M_{\rm mess} \sim |\lambda_\Phi Y| \,\sim \, \frac{\lambda_\Phi
A_\kappa}{6\sqrt{3 D_A}}v_{\rm PQ},\eea
and the soft scalar masses are \bea
 m_{\varphi_i}({\rm MM}) &\sim & \frac{m_{3/2}}{2a\tau_s},
\quad
m_{\varphi_i}({\rm D}) \sim  \sqrt{D_A},
\quad
 m_{\varphi_i}({\rm GM}) \sim
\frac{6\sqrt{D_A}}{8\pi^2 A_\kappa} \sqrt{D_A},
\eea 
where  MM $=$ moduli mediation mainly due to $T_s$ dependence of $Z_i$ in Eq.~(\ref{mat_action}), and of the order of $m_{3/2}/2a\tau_s= m_{3/2}/8\pi^2$ \cite{lvs}, 
D $=$ $D$-term contribution induced by the $U(1)_A$ \cite{Choi:2010gm}, GM $=$ gauge mediation (\ref{gauge}). 
Since $A_\kappa \sim \sqrt{D_A}/8\pi^2$ and all dimensionless parameters are of order unity, the above results give the following qualitative pattern of mass scales:
\bea 
&& v_{\rm PQ} \sim \sqrt{m_{\rm soft} M_{\rm GUT}},
\nonumber\\
&& M_{\rm mess} \,\sim\, \frac{v_{\rm PQ}}{16\pi^2}, \nonumber\\
&&m_{\rm soft}({\rm D})\sim m_{\rm soft}({\rm GM}) 
\sim 8\pi^2 m_{\rm soft}({\rm MM}) \,\sim \,  
m_{3/2}.
\eea 
We then have enough parameters to make gauge mediation dominates soft terms and all squarks and sleptons have positive masses squared. 
Since $Z$ and $S$ are neutral under the $U(1)_A$, 
in this case, the dominant contribution to the soft mass-squared of $Z$ and $S$ in LVS is 
from modulus mediation. Hence at UV input scale, we expect
\beq
m_Z \sim m_S \sim m_{\rm soft}({\rm MM})\sim \frac{m_X}{8\pi^2}
\eeq
At low energy scale, if $\lambda_S = \mathcal{O}(1)$, RG-running of the soft mass-squared of $Z$ becomes strong.
As the result, potential around the origin along $Z$ direction can have tachyonic instability, and $Z$ develops intermediate scale VEV which reproduce $\mu$-term
\footnote{
$S$ could also have a similar instability while $Z$ is held around the origin.
However, the RG-running of $m_S^2$ caused by the $\lambda_S$ interaction term is slower than that of $m_Z^2$, so $Z$ can develop large VEV, providing large mass to $S$, while $S$ is still at around the origin.
As the result, $S$ can not be destabilized.  
}. 

Since the gravitino mass $m_{3/2}$ is around soft mass scale in the visible sector, 
the large volume modulus mass will be around 
\bea
m_{\phi_b} \sim 0.01\GeV - 1\GeV.
\eea
Thus, the one-step thermal inflation is not enough in this region.
However, we noticed that singlet scalars in the model \eq{W-DTI} can have a natural hierarchy in their soft mass-squared so that 
\beq
m_X \gg m_Z, \quad X_0 \gg Z_0
\eeq
and hence
\beq 
T_{{\rm c},X} \gg T_{{\rm c},Z} , \quad V_{X,0} \gg V_{Z,0}
\eeq
where $V_{Z,0} \equiv |V_Z(Z_0)|$ and $V_{X,0} \equiv V_0 - V_{Z,0}$. 
Therefore, a double-stage thermal inflation can be realized.
Additionally, a late-time Affleck-Dine leptogenesis may have a chance to work.
$LH_u$ flat direction is also neutral under $U(1)_A$, so it does not have tree-level $D$-term mass associated with the $U(1)_A$
\footnote{
There is a dominant negative 1-loop contribution \cite{Shin:2011uk} which drives the soft mass-squared of $LH_u$ be positive around the electroweak scale.
If this is all, $LH_u$ flat direction would be stable around the origin even without $\mu$-term contribution.
}.
However, the presence of gauge mediation effect from $F_Y$ provides $LH_u$ a soft scale mass-squared which is expected to RG-run to a negative value at low energy scale.
Around electroweak scale, the absolute value of the mass-squared is typically larger than that of the singlet $Z$, and note that gauge mediation effect on $LH_u$ takes place only after $X$ is destabilized.
%
Therefore, we expect the following order of destabilization.
\beq
T_{{\rm c,} X} > T_{{\rm c}, LH_u} > T_{{\rm c}, Z}
\eeq
$LH_u$ is eventually lifted up and stabilized around the origin as $Z$ develops intermediate scale VEV reproducing $\mu$-term.
This is a perfect circumstance to realize the late-time Affleck-Dine leptogenesis \cite{Stewart:1996ai,Jeong:2004hy,Felder:2007iz,Kim:2008yu,Choi:2009qd}.
%

In the following subsections we will describe how the cosmological problem of the light modulus can be solved or ameliorated by the double thermal inflation appearing in the model \eq{W-DTI}.
We also check if right amounts of baryon asymmetry and dark matter can be obtained for the parameter set solving moduli problem.

\subsection{Dilution of moduli by double thermal inflation}

In our model (\eq{W-DTI}), all the directions including $X$, $Y$, $Z$ and $LH_u$ flat directions, but except moduli, are expected to be held around origin at high temperature after primordial inflation.
The first thermal inflation begins when $\rho_{\phi_b} \sim V_0 \sim V_{X,0}$ at a temperature
\beq
T_1 \sim \rho_{\phi_b, 0}^{1/4} \left( \frac{H_X}{m_{\phi_b}} \right)^{2/3} \sim \rho_{\phi_b, 0}^{-1/12} V_{X,0}^{1/3}
\eeq
with $H_X \equiv V_{X,0}^{1/2} / \left( \sqrt{3} M_\planck \right)$, and ends when temperature drops to $T_{{\rm c},X}$ and $X$ is destabilized from the origin.
Subsequently, $LH_u$ flat direction is destabilized.
Both of $X$ and $LH_u$ condensations contribute to the radiation density by their partial decays. 
The second thermal inflation can occur if the background temperature is higher than $T_{{\rm c},Z}$ when $H \sim H_Z \equiv V_{Z,0}^{1/2} / \left( \sqrt{3} M_\planck \right)$.
The radiation energy density of standard model particles at the time is 
\bea \label{rhor-at-t2}
\rho_{\rm r} 
&\simeq& \Delta \rho_{{\rm r}, LH_u} + \Delta \rho_{{\rm r}, X}
\nonumber \\
&\simeq& \sqrt{3} \mathrm{Br}(X \to {\rm SM}) \Gamma_X M_\planck V_{Z,0}^{1/2} \left[ 1 + \frac{\Gamma_{LH_u \to \rm SM}}{\Gamma_{X \to \rm SM}} \frac{\rho_{{\rm r}, LH_u}(t_{\rm c, 1})}{\rho_{{\rm r}, X}(t_{\rm c, 1})} \right]
\nonumber \\
&=& \sqrt{3} \mathrm{Br}(X \to {\rm SM}) \Gamma_X M_\planck V_{Z,0}^{1/2} \left[ 1 + \frac{8\,\gamma_{LH_u}}{\mathrm{Br}(X\to {\rm SM})}   \left( \frac{m_{LH_u}}{m_X} \right)^5 \right]
\eea
where $\Gamma_X = \Gamma_{X \to \rm SM} + \Gamma_{X \to aa}$ is the total decay rate of $X$ with  
\bea
\Gamma_{X \to \rm SM} &=& \frac{1}{8 \pi} \gamma_X \left( \frac{\alpha_s}{4 \pi} \right)^2 \frac{m_X^3}{X_0^2}
\\
\Gamma_{X \to a a} &=& \frac{1}{64 \pi} \frac{m_X^3}{X_0^2},
\eea
and $\gamma_X = \mathcal{O}(1)$,  and
\bea
\Gamma_{LH_u \to \rm SM} &=& \frac{1}{8 \pi} \gamma_{LH_u} 
\frac{ m_{LH_u}^3}{|\ell_0|^2}
\eea
with $\gamma_{LH_u} = \mathcal{O}(1)$ and $m_{LH_u}\sim m_{\rm soft}/\sqrt{8\pi^2}$ being the effective mass for the $LH_u$ directional field at $|\ell_0|$, which is originated from the fact that $LH_u$ is stabilized by radiative effect rather than tree-level seesaw operator.
In the last line of \eq{rhor-at-t2} we used 
\bea
\rho_{{\rm r}, LH_u}(t_{\rm c, 1}) &\sim& m_{LH_u}^2 |\ell_0|^2
\\
\rho_{{\rm r}, X}(t_{\rm c, 1}) &\sim& m_X^2 X_0^2
\eea
For intermediate scale of $X_0$ and $Z_0$ with $m_X \sim \left( 8 \pi^2 \right) m_Z$ and $m_Z \sim T_{{\rm c}, Z}$, one finds $\rho_{\rm r} \gg T_{{\rm c},Z}^4$, hence the second thermal inflation can take place.
Note that the contribution from $LH_u$ flat direction to radiation after the first thermal inflation is larger than that from $X$ for rather mild hierarchy between $m_{LH_u}$ and $m_{X}$, so the additional dilution from $LH_u$ decay can be obtained with a factor
\beq
\Delta_{LH_u} 
\equiv \left( \frac{\Delta \rho_{{\rm r}, LH_u}(t_{\rm c,2})}{\Delta \rho_{{\rm r}, X}(t_{\rm c,2})} \right)^{3/4} 
= \left[ \frac{8 \gamma_{LH_u}}{{\rm Br}(X \to {\rm SM})} \left( \frac{m_{LH_u}}{m_X} \right)^5 \right]^{3/4}
\eeq

Soon after the second thermal inflation, the energy density of the universe is nearly equally distributed to the radial and axial components of $Z$.
Hence eventual reheating temperature is determined by the one which decays later than the other.
For $m_Z = \mathcal{O}(10) \GeV$, axial component decays later with a rate given by \eq{mu-term-decay} with $m_{a_X}$ replaced to $m_{a_Z}$ (the mass of axial component of $Z$).
Combined with the effect of the partial decay of $X$, the significant amount of entropy release in this decay provide a huge dilution to moduli.

\begin{figure}[ht] 
\centering
\includegraphics[width=0.6\textwidth]{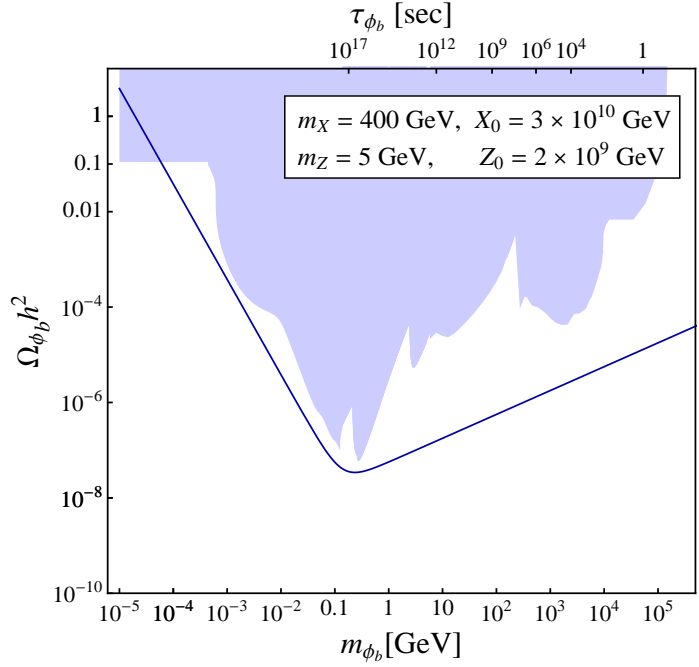} 
\caption{The total abundance of moduli at decay under the double-stage of thermal inflation realized in the model, \eq{W-DTI}, with constraints given by Fig.~\ref{fig:mod_problem}.
Parameters used are $\tan \beta = 10$, $m_{\rm soft} = 1 \TeV$, $B = 0.5 m_{\rm soft}$, $\mu = m_A = m_{\rm soft}$, $m_X =400 \GeV$, $m_Z = m_{a_Z}=5\GeV$, $T_{{\rm c},X} = m_X$, $T_{{\rm c},Z} = m_Z$, $X_0 = 3\times 10^{10}\GeV$ and $Z_0 = 2\times 10^9\GeV$.
They are taken to provide correct relic densites of  baryon and dark matter 
(Fig.~\ref{fig:bau-cdm-mod}).}
\label{fig:moduli-in-DTI}
\end{figure}

In Fig.~\ref{fig:moduli-in-DTI}, we show 
the late-time total abundance of moduli after double thermal inflation.
We use $m_X=400 \GeV$, $m_Z = m_{a_Z}= 5 \GeV\sim m_X / \left( 8 \pi^2 \right) 
$ and $T_{\rm c, i} = m_i$ reflecting the constraints from baryon number and dark matter density
studied in next subsection. 
The figure shows that the whole range of $m_{\phi_b}$ can be cosmologically viable. 
Therefore, a double-stage thermal inflation provides a complete solution to the cosmological moduli problem of the light volume modulus in LVS.
%
%

For the parameters we have used in Fig.~\ref{fig:moduli-in-DTI}, we found $N_{e, \rm tot} \sim 17$.
Although such an amount of additional $e$-foldings does not touch CMB scale, some of inflation model is incompatible with this \cite{Kawasaki:2009hp}.

\subsection{Baryogenesis}
In our model, $LH_u$ flat-direction is destabilized after the end of the first thermal inflation, but before the second thermal inflation.
This make the late-time Affleck-Dine leptogenesis work as $Z$ is destabilized and reproduces $\mu$-term \cite{Jeong:2004hy}.

Including the dilution due to entropy release in the eventual decay of $Z$, the resulting baryon asymmetry at present is estimated as \cite{Jeong:2004hy}
\beq \label{YBpresent}
\frac{n_B}{s} \sim \frac{n_B}{n_Z} \frac{T_Z}{m_Z} \sim \frac{n_L}{n_\mathrm{AD}} \frac{n_\mathrm{AD}}{n_Z} \frac{T_Z}{m_Z} \sim \frac{n_L}{n_\mathrm{AD}} \frac{m_{LH_u}}{m_Z} \left( \frac{|l_0|}{Z_0} \right)^2 \frac{T_Z}{m_Z}
\eeq
where $n_Z$, $n_L$ and $n_\mathrm{AD}$ are number densities of $|Z|$, lepton asymmetry and AD field, respectively, $m_{LH_u}$ is the mass scale of $LH_u$, and $|\ell_0|$ is the field-value of $LH_u$ when it is lifted up as $Z$ reaches its VEV.
For a small $CP$-violating phase, $\delta\ll 1$, the conserved lepton asymmetry can be expressed as 
\beq
n_L = \alpha \, \delta \, m_\theta |\ell_0|^2
\eeq
where $\alpha \sim 0.1$ is the efficiency factor of conserving the generated asymmetry \cite{Felder:2007iz,Kim:2008yu}, and $m_\theta$ is the mass of the angular mode of the $LH_u$ direction when it is lifted up and starts to roll in.
We find
\beq
m_\theta^2 \sim \mu \ \frac{\lambda^2_L |l_0|^2}{\lambda_N Z_0} \sim \frac{m_\nu \mu}{v^2 \sin^2 \beta} |\ell_0|^2 \simeq 3 \GeV^2 \left( \frac{m_\nu}{0.1 \eV} \right) \left( \frac{\mu}{1 \TeV} \right) \left| \frac{\ell_0}{10^6 \GeV} \right|^2
\eeq
where $m_\nu$ is the mass of a left-handed neutrino, $v = 174 \GeV$ is the VEV of neutral Higgs and we took $\tan \beta = 10$.
Hence   
\beq
\frac{n_L}{n_\mathrm{AD}} 
\sim \alpha \, \delta \, \left( \frac{m_\theta}{m_{LH_u}} \right) 
= 10^{-4} \left( \frac{\alpha}{0.1} \right) \left( \frac{\delta}{0.1} \right) \left( \frac{m_\theta}{2 \GeV} \right) \left( \frac{200 \GeV}{m_{LH_u}} \right)
\eeq
and 
\beq \label{nBs}
\frac{n_B}{s} \sim 2 \times 10^{-10} \left( \frac{n_L / n_\mathrm{AD}}{10^{-4}} \right) \left( \frac{m_{LH_u}/m_Z}{20} \right) \left( \frac{|\ell_0| / Z_0}{10^{-3}} \right)^2 \left( \frac{T_Z}{1 \GeV} \right) \left( \frac{10 \GeV}{m_Z} \right)
\eeq
Therefore, the obtained baryon asymmetry may match the observation within the uncertainties of involved parameters.

\subsection{Dark matter}
The dark matter candidates in our scenario are QCD-axion and the LSP which is flatino, the fermionic superpartner of $Z$,  whose mass is expected to be $\mathcal{O}(10) \GeV$.
For $v_{\rm PQ} \sim 10^{10}-10^{11} \GeV$ as the PQ-scale in our scenario, the relic density of QCD-axion is subdominant.
On the other hand, abundant flatinos can be produced in the decays of the next-lightest-supersymmetric particle (NLSP) denoted as $\chi$ which might be neutralino.
For the parameter set giving a right amount of baryon asymmetry, if kinematically allowed, the decay of $Z$ produces too much flatinos, hence it should be forbidden \cite{Kim:2008yu} and it is the case since $m_{{\tilde a}_Z} \approx m_Z$ is expected.
However, production from decays of NLSP can provide a right amount of flatinos to match observation.
The decay rate of the (neutralino) NLSP to flatino is given by  
\beq
\Gamma_\nlsp \simeq \frac{1 }{8 \pi}\gamma_\nlsp \frac{m_\nlsp^3}{Z_0^2} 
\eeq
with $\gamma_\nlsp = \mathcal{O}(1)$.
The present abundance of axino dark matter for $T_Z \ll T_\nlsp \equiv m_\chi / 10$ is then \cite{Kim:2008yu} 
\beq \label{approxomeganlsp} \hskip -0.29cm
\Omega_\axino  h^2 \sim \mathinner{0.08} \gamma_\nlsp
\left( \frac{10^3 g_*^{3/2}(T_Z)}{g_*^3(T_\nlsp)} \right)
\left( \frac{m_\nlsp}{10^2 \GeV} \right)
\left( \frac{m_\axino}{1 \GeV} \right)
\left( \frac{10^{9} \GeV}{Z_0} \right)^2
\left( \frac{10^2\, T_Z}{m_\nlsp} \right)^7
\eeq
We find that for $m_\chi \sim 200 \GeV$, $m_\axino \sim m_Z \sim 5 \GeV$, $Z_0 = \mathcal{O}(10^9) \GeV$ with $T_Z = \mathcal{O}(1) \GeV$ can match the observed dark matter relic density, $\Omega_\axino h^2 \simeq 0.11$.

\begin{figure}[h] 
\centering
\includegraphics[width=0.6\textwidth]{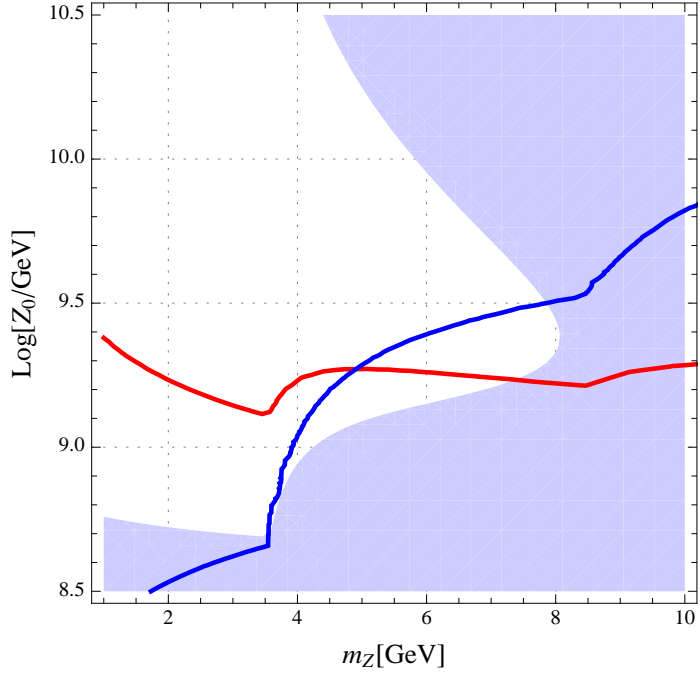} 
\caption{Contours corresponding to the observed baryon number asymmetry (red) and dark matter (blue) as functions of  $m_Z$ and $Z_0$ for $m_{\phi_b} = 0.08 \GeV$, $m_\chi = 150 \GeV$ and $n_L / n_{\rm AD} = 10^{-4}$, $m_{LH_u} = 200 \GeV$, $\ell_0 = 10^6 \GeV$, $g_*(T_Z) = g_*(T_\chi) = 100$ with all the other parameters being the same as Fig.~\ref{fig:moduli-in-DTI}.
The light blue colored region is excluded due to moduli over-production.
In the region above the red/blue line, baryon number asymmetry and dark matter at present are smaller than observed ones.
  }
\label{fig:bau-cdm-mod}
\end{figure}

The abundance of moduli, baryon number asymmetry and relic density of dark matter depend crucially and commonly on $m_Z$ and $Z_0$, hence in Fig.~\ref{fig:bau-cdm-mod} we show contours corresponding to the observed baryon number asymmetry (red) and dark matter (blue)  with the bound on moduli abundance (the boundary of the colored region) in ($m_Z$, $Z_0$) plane for $m_X = 400\GeV$.
In the figure, the light blue colored region is excluded due to moduli over-production.
As shown in the figure, right amounts of baryon number asymmetry and dark matter can be obtained simultaneously for $m_{\phi_b} \lesssim 0.1 \GeV$.


\section{Conclusion}
In this paper, we examined the cosmological moduli problem in large volume scenario in which the overall volume modulus
$\tau_b$  is stabilized at a value of $\mathcal{O}(10^4)$ to generate
 the GUT to the Planck scale ratio $M_{\rm GUT}/M_{\rm P}\sim 1/\tau_b^{1/2}\sim 10^{-2}$.

We found  that if the primordial inflaton decay rate $\Gamma_I \gtrsim m_{\tau_b}$, single thermal inflation can solve the cosmological moduli problem only for a limited range of $m_{\tau_b}$.
Particularly, the mass range $m_{\tau_b} = \mathcal{O}(10^{-2} - 1) \GeV$ is viable only if the decay temperature of flaton for thermal inflation is about few $\MeV$, for which baryogenesis is difficult to be implemented.
On the other hand, if primordial inflatons decay just before thermal inflation begins,  the whole relevant modulus mass range in consideration, i.e. $m_{\tau_b}=\mathcal{O}(10^{-2}-10^5)$ GeV,  can be viable with single thermal inflation, while  allowing a successful late-time Affleck-Dine leptogenesis.

Since such a late decay of primordial inflatons is not typical, and may not be realized in nature, we 
examined an alternative possibility,  double thermal inflation
as a complete solution of the cosmological moduli problem in large volume scenario.  
Considering a concrete example, we showed that double thermal inflation can be realized in large volume scenario in a natural manner,  and the cosmological problem of the light volume modulus can be solved for the whole relevant mass range.
We also showed that late-time Affleck-Dine leptogenesis can work after the second thermal inflation, and  flatino LSPs with a mass of few $\GeV$, which are produced through the decays of the visible sector NLSP (e.g. neutralino), can provide a right amount of dark matter at present.

\section*{Acknowledgement}

KC were supported by the National Research Foundation of Korea (NRF) grant (No. 2007-0093865 and No. 2012R1A2A2A05003214) and the BK21 project funded by the Korean Government (MEST). 
WIP is supported in part by Basic Science Research Program through the National Research Foundation of Korea(NRF) funded by the Ministry of Education, Science and Technology(2012-0003102).
CSS were supported by Basic Science Research Program through the National Research Foundation of Korea (NRF) funded by the Ministry of Education, Science and Technology (No. 2011-0011083). 
CSS acknowledges the Max Planck Society (MPG), the Korea Ministry of
Education, Science and Technology (MEST), Gyeongsangbuk-Do and Pohang
City for the support of the Independent Junior Research Group at the Asia Pacific
Center for Theoretical Physics (APCTP).

\end{document}